%% file: 0-puff.tex
\documentclass[twocolumn]{article}
\usepackage[utf8]{inputenc}
\usepackage[T1]{fontenc}
\usepackage[UKenglish]{babel}
\usepackage[english]{varioref}
\usepackage[nottoc]{tocbibind}
\usepackage[toc,page]{appendix}
\usepackage{caption}
\usepackage{amsmath}
\usepackage{amsthm}
\usepackage{amssymb}

\usepackage{mathrsfs}
\usepackage{geometry}
\usepackage{float}
\usepackage[caption = false]{subfig}

\usepackage{graphicx}
\usepackage{array}

\usepackage{hyperref}
\hypersetup{
    colorlinks = True,
    linkcolor = blue,
    citecolor = blue
}
\usepackage[font = small,labelfont = bf]{caption}
\usepackage{booktabs}

\usepackage{csquotes}

\usepackage{authblk}
\usepackage{subfiles}

\usepackage[table, svgnames, dvipsnames, x11names]{xcolor}

\usepackage{tikz}
\newcommand{\coloredsquare}[2][black]{%
  \tikz[scale=0.7, baseline=(X.base)]{
    \node (X) at (0,0) {}; 
    \draw[line width=0.4pt, fill=#2, draw=#1] (0,0) rectangle (0.4cm,0.4cm);
  }%
}

\definecolor{MyGreenish}{RGB}{150,210,104}
\definecolor{MyLightGreenYellow}{RGB}{215, 241, 145}
\definecolor{MyYellowOrange}{RGB}{253, 202, 121}
\definecolor{MyReddishOrange}{RGB}{246, 122, 73}
\definecolor{Silver}{RGB}{220,220,220}

\usepackage{multirow}
\usepackage{rotating}
\usepackage{microtype}

\usepackage{lmodern}

\usepackage{titlesec}
\titleformat*{\section}{\bfseries}
\titleformat*{\subsection}{\itshape}
\titleformat*{\subsubsection}{\itshape}

\newcommand{\MWsqm}{\text{MW} \times \text{m}^{-2}}

\newcommand{\els}{\text{el} \times \text{s}^{-1}}
\newcommand{\ats}{\text{at} \times \text{s}^{-1}}
\newcommand{\mcube}{\text{m}^{3}}
\newcommand{\permcube}{\text{m}^{-3}}

\newcommand{\GammaD}{\Gamma_{\text{D}}}
\newcommand{\GammaX}{\Gamma_{\text{Z}}}
\newcommand{\fQZ}{\gamma_{\text{DZ}}}
\newcommand{\VOFL}{V_{\text{div}}}
\newcommand{\SOFL}{S_{\text{div}}}
\newcommand{\VCFL}{V_{\text{p}}}
\newcommand{\Ldiv}{L_{\text{div}}}
\newcommand{\Rdiv}{R_{\text{div}}}
\newcommand{\Hdiv}{H_{\text{div}}}
\newcommand{\Ndiv}{N_{\text{div}}}
\newcommand{\pdiv}{p_{\text{div}}}
\newcommand{\Tesep}{T_{\text{sep}}}
\newcommand{\nesep}{n_{\text{sep}}}

\newcommand{\neGW}{n^{\text{G}}}
\newcommand{\neSudo}{n^{\text{Su}}}
\newcommand{\Psep}{P_{\text{sep}}}
\newcommand{\Bpol}{B_{\text{pol}}}
\newcommand{\lqEich}{\lambda_q^{\text{E}}}
\newcommand{\Lconn}{L_{\text{conn}}}
\newcommand{\fLH}{f^{\text{S}}}
\newcommand{\Seff}{S_{\text{eff}}}

\title{\textbf{Multi-Machine Scaling Laws for Fuel and Impurity Puffing Rates Sufficient for Detachment Access: a Systematic Review of Magnetic Confinement Fusion Devices}}

\author[1]{M. Moscheni*}
\author[1]{A. Herrmann}
\author[1]{R. Kembleton}
\author[2]{M. Kryjak}
\author[1]{S. Lazerson}
\author[1]{F. Levi}
\author[3,4]{M. Siccinio}
\author[1]{P. Staniec}
\author[5]{T. Giegerich}
\author[5]{C. Tantos}
\author[1]{the Gauss Fusion Team}

\affil[1]{Gauss Fusion GmbH, Parkring 29, 85748 Garching bei M\"unchen, Germany}
\affil[2]{United Kingdom Atomic Energy Authority, Culham Campus, Abingdon, OX14 3DB, United
Kingdom}
\affil[3]{Max-Planck-Institut f\"ur Plasmaphysik, 85748 Garching bei M\"unchen, Germany}
\affil[4]{EUROfusion Consortium, 85748 Garching bei M\"unchen, Germany}
\affil[5]{Karlsruhe Institute of Technology, Hermann-von-Helmholtz-Platz 1, 76344 Eggenstein-Leopoldshafen, Germany}

\begin{document}

\maketitle

\begin{abstract}

    An open-source database of 457 experimental and numerical entries representing 32 machines---including tokamaks, stellarators, and linear plasma devices---is assembled. From this dataset, we derive multi-machine scaling laws that predict the fuel and impurity puffing rates sufficient to edge plasma detachment---the leading reactor-relevant solution to the challenge of plasma-wall interaction. Validation against up to 40 L- and H-mode plasmas shows agreement within a factor of 1.5 in about 50\% of cases, and within a factor of 2 on average. Divertor volume alone is found to strongly correlate with the fuelling rate. Inclusion of plasma opaqueness leads to $\GammaD \propto [\nesep\, a\, (S_{\text{div}}/V_{\text{div}})^{-1.5}]^{1.05}$, valid across all toroidal devices. Its H-mode simplification, $\GammaD^{\text{HDL}} \propto 0.43\, a^{1.58}\, \lambda_q^{-0.89}$, avoids explicit dependence on $\nesep$ and carries intrinsic physical meaning through the H/L density limit and the power fall-off length. The impurity seeding rate is captured by a general non-linear law, from which the Greenwald-Eich-Scarabosio simplification, $\GammaX^{\text{GES}} \propto a^{1.51}\, \lambda_q^{-0.27}$, is obtained. Similar relationships are defined for stellarators, consistent with tokamak trends but still awaiting validation---an opportunity for further study. These results have immediate relevance for reactor fuel-cycle design and edge plasma modelling. More broadly, they demonstrate that physics-based 0D laws can reliably link detachment access to engineering actuators, offering practical tools for reactor design. Our laws represent macroscopic patterns across machines rather than microscopic variations within an individual device---providing the basis for our forthcoming studies aimed at extending this framework to machine-specific behaviour.

\end{abstract}

{\small*E-mail: matteo.moscheni@gauss-fusion.com}


    \section{Introduction}\label{sec:introduction}
    
        The past few decades witnessed a speciation of magnetic confinement fusion devices culminating in the coexistence of designs for conventional \cite{Federici_2019_EU_DEMO}, spherical \cite{Meyer_STEP}, high-field tokamaks \cite{SORBOM2015378_ARC} and stellarators \cite{WARMER2024114386_stellarators} as possible candidates to harness nuclear fusion.
        
        Common denominator is the open issue of the edge plasma and its interactions with the divertor plasma-facing components \cite{donne2019, Krieger_2025, stangeby2000plasma, Zohm_2013}---where the edge plasma must be tuned such that realistic engineering can simultaneously satisfy tight constraints of power and particle exhaust (few tens of $\MWsqm$ and hundreds of $\text{Pa} \times \mcube \times \text{s}^{-1}$, respectively \cite{PITTS201760_engineering, Henderson_2025}), and divertor survival ($\sim 5$ year lifetime \cite{Federici_2017_materials}). The most widely accepted solution is the detached edge plasma scenario \cite{Krasheninnikov_Kukushkin_2017, Soukhanovskii_2017}, a highly-dissipative state that cools the plasma to protect the divertor materials, and which is established by puffing gaseous fuel and seeded impurities---focus of the present study. In pilot plants, these injection rates will likely be controlled in real-time to maintain the detached state \cite{Henderson_2023, Yang_2024, Park_2024}, whereby keeping the plasma-divertor interaction suppressed during the plasma discharge.

        Quantifying the required fuel puffing rate and impurity seeding ratio presents one of the most significant challenges in the design of the tritium fuel cycle (TFC \cite{Abdou_2021, Meschini_2023}) associated to a magnetic confinement fusion device. Designing a TFC with an under-estimated puffing rate would preclude a safe detached operation altogether. However, over-estimating for a "worst case scenario" with the highest possible feed rate requirements and later relying on a reduced operation of the TFC is also impractical. This would lead to an oversized plant, requiring unnecessary extra tritium inventory (with regulator and safety considerations), as well as incurring additional capital and operational costs. It may also lead to sub-optimal technology selections. Thus, potential puffing values spanning several orders of magnitude can not be realistically used as a starting point for a reactor design and a more reliable foundation must be established.
    
        The task of estimating puffing rates compatible with a robust detached operation is traditionally tackled via state-of-the-art suites dedicated to 2D edge plasma simulations (e.g. SOLPS-ITER \cite{XavierBONNIN2016, kotov_reiter_kukushkin_bochum}, SOLEDGE3X \cite{Rivals_Soledge3x} and EDGE2D-EIRENE \cite{Simonini_1994, Reiter2019EIRENE}), uniquely predisposed to model the multi-species transport from the puffing injection valve to the divertor. Multi-parameter scans with such tools ultimately bound, among many others, the admissible puffing rates for fuel and impurities \cite{Osawa_2024, RUBINO2021100895, BALBINOT2021100952, Islam_2024}.
        
        Comprising in excess of 125 SOLPS-ITER simulations for a single operating point---each worth up to a month of wall clock time \cite{Lore_2022}, the ITER edge plasma database \cite{PITTS2019100696} is among the most comprehensive ones ever assembled, and is yet constantly expanded \cite{Lore_2022}. Despite its recognised success, this approach is hardly replicable for other tokamaks, notwithstanding the rapid pace pick-up of stellarator designs \cite{Bader, LION2025114868} and their need of 3D simulations \cite{Winters_2021}.

        Other authors acknowledged the above. Some distilled regression laws---but specific to a single device, not accounting for detachment \cite{Saarelma_2018, Luda_2020} or impurities \cite{Kukushkin_2003, PACHER2015591}. Others created promising neural network surrogates \cite{DASBACH2023101396, Wiesen_2024}, though at the cost of losing physical interpretability and, for the time being, only applicable for certain configurations. And still others successfully constructed 0.5D physical models \cite{body2025, Kallenbach_2016, Henderson_2025, Siccinio_2016, Kallenbach_2018, Cowley_2022, Silvagni2025Scaling, Goldston_2017} which provide valuable insight but, nonetheless, necessitate of assumptions, estimates and understanding beyond those needed for 0D scaling laws. 

        Here we follow the recommendation of Lore \textit{et al.} \cite{Lore_2022}: ``[...] the relationship between the actuators and effective sorting variables should be explored''---but we apply it at large. Complementing the afore-mentioned studies, the present work scrutinises the puffing rates across devices via a human-driven systematic review of published data for stably detached edge plasmas to obtain widely-applicable 0D laws.

        The paper is organised as follows. Section \ref{sec:methods} illustrates the features of the database employed in the analyses, the conventions adopted and the assessment strategy. Section \ref{sec:results} presents the results of the study---scaling law derivation and validation. Section \ref{sec:discussion} discusses this work compared to the existing knowledge, its interpretation and caveats. Practical examples applied to reactor concepts are also provided, before drawing the conclusions in section \ref{sec:conclusion}.

    \section{Methods}\label{sec:methods}

        \subsection{Database overview}\label{sec:methods_database}
    
            The database here considered (available at \cite{zenodo_repo}) groups experimental and numerical data from 20 conventional tokamaks (``CTKs'', e.g. HL-2A \cite{Gao_2023}, JFT-2M \cite{Kawashima_1999}), 5 spherical tokamaks (``STKs'', e.g. NSTX \cite{MEIER20151200}), 2 high-field tokamaks (``HFTKs'', e.g. SPARC \cite{Lore_2024}) and 3 stellarators (``STLs'', e.g. LHD \cite{MUKAI2022101294}). With the addition of 5 linear plasma devices (``LPDs'', e.g. Magnum-PSI \cite{Tanaka_2020}, MAP-II \cite{Okamoto_2006}), all the magnetic confinement fusion devices are encompassed. Machine IDs and references are reported in Appendix \ref{apx:nomenclature}.

            The salient features of the database are summarised in table \ref{tab:methods_database} and further detailed in section \ref{sec:methods_variables}. Overall, 5 orders of magnitude worth of machine volume, and 3 of power, are spanned throughout the 32 different devices expounded in 124 scientific publications. Each of the 193 entries from 7 different simulation tools \cite{XavierBONNIN2016, Rivals_Soledge3x, Simonini_1994, BUFFERAND2017852, ROGNLIEN1992347, kawashima2006development, Zagorski_2008} and each of the 264 experimental instances correspond to edge plasmas exhibiting stable detachment at the outer target without MARFE \cite{Lipschultz1984Marfe} (section \ref{sec:methods_conventions_puffing_detachment}). The database as a whole encompasses a wide range of features, including divertor geometry (open or closed divertor \cite{FEVRIER2021100977}), magnetic topology (whether linear \cite{Hayashi_2016}, diverted double-null with \cite{HAVLICKOVA20151209} or single-null without \cite{Komm_2019} advanced magnetic configurations \cite{Reimerdes_2020, MILITELLO2021100908, Lunt_PRL_CRD}, or island divertors \cite{Effenberg_2019}), seeded impurity species \cite{ZAGORSKI201637}, divertor material (e.g. molybdenum \cite{LIPSCHULTZ1997771}, carbon \cite{Chen_2018}), type of radiation emission (e.g. traditional or X-point radiator \cite{Senichenkov_2021}) and puff location \cite{Osawa_2024, ZHOU2022113222}. Last but not least, transient and steady L-modes \cite{GUILLEMAUT2013S638} and H-modes (EDA \cite{LaBombard_2011} and not \cite{Lan_2020}) populate the database.

            \begin{table*}[htbp]
            \centering
            \renewcommand*\arraystretch{1.05}
            \begin{tabular}{lccccc}
                \toprule
                \multirow{3}{*}{Variable} & \multirow{3}{*}{Units} & \multicolumn{2}{c}{Training set} & \multicolumn{2}{c}{Validation set} \\
                & & \multicolumn{2}{c}{{(DOD$>$1)}} & \multicolumn{2}{c}{{(DOD=1)}} \\
                & & Min. & Max. & Min. & Max. \\\midrule
                n. data & -- & -- & 412 & 16 & 45 \\
                $R_0$ & $\text{m}$ & 0.36 & 9 & 0.7 & 8.9 \\
                $a$ & $\text{m}$ & 0.16 & 2.9 & 0.25 & 2.9 \\
                $A$ & -- & 1.2 & 12 & 1.4 & 6.2 \\
                $\VCFL$ & $\mcube$ & 6.5E$-$01 & 2.5E+03 & 1.6E+00 & 2.5E+03 \\
                $\VOFL$ & $\mcube$ & 3.8E$-$03 & 4.8E+02 & 6.9E$-$02 & 1.5E+02 \\\midrule
                n. L-modes & -- & -- & 80 & -- & 20 \\
                n. H-modes & -- & -- & 294 & -- & 21 \\
                $\Psep$ & $\text{MW}$ & 1.2E$-$01 & 2.0E+02 & 2.1E$-$01 & 1.7E+02 \\
                $\lqEich \times \fLH$ & $\text{mm}$ & 0.25 & 14 & 0.25 & 27 \\
                $\neGW$ & $\permcube$ & 3.0E+19 & 8.5E+20 & 3.0E+19 & 6.7E+20 \\
                $\neSudo$ & $\permcube$ & 4.3E+19 & 1.5E+20 & -- & -- \\
                $\nesep$ & $\permcube$ & 3.5E+18 & 2.0E+20 & 3.3E+18 & 1.2E+20 \\
                $\nesep \times a$ & $\text{m}^{-2}$ & 2.0E+18 & 2.0E+20 & 1.7E+18 & 1.2E+20 \\\midrule
                $\GammaD$ & $\els$ & 4.2E+19 & 2.5E+24 & 1.0E+20 & 5.0E+23 \\
                $\GammaX$ & $\els$ & 2.2E+19 & 4.9E+23 & 2.2E+19 & 2.1E+23 \\
                $\GammaD / \GammaX$ & -- & 4.1E$-$03 & 3.9E+02 & 9.5E$-$03 & 4.6E+01 \\\bottomrule
            \end{tabular}
            \caption{Main variables accompanied by their units and their minimum (Min.) and maximum (Max.) values in the database for the cases past (DOD$>$1, training set) and at (DOD=1, validation set) the detachment onset (see text for details).}
            \label{tab:methods_database}
        \end{table*}

        \subsection{Variables and conventions}\label{sec:methods_variables}

            \subsubsection{Geometrical parameters}\label{sec:methods_conventions_geometry}

                Nominal values of geometrical parameters (e.g. plasma elongation $\kappa$) are added where lacking in the database. The variables considered and listed in table \ref{tab:methods_database} are: the machine major and minor radii, $R_0$ and $a$, respectively; the aspect ratio $A = R_0 / a$; the plasma and divertor volumes, $\VCFL$ and $\VOFL$, respectively. The plasma volume $\VCFL$ is computed as $2 \pi^2 R_0 a^2 \kappa$. The volume ($\VOFL$) and surface ($\SOFL$) of the divertor are approximated by simplistic generalisations:
                
                \begin{equation}\label{eq:VOFL_SOFL}
                    \begin{split}
                        \VOFL & = \Ndiv \times f_{\text{tor}} \times 2 \pi \Rdiv \times (\Ldiv \times \Hdiv) \; , \\
                        \SOFL & = \Ndiv \times f_{\text{tor}} \times 2 \pi \Rdiv \times (\Ldiv + \Hdiv) \times 2 \; .
                    \end{split}
                \end{equation}
                
\begin{figure}
                    \centering
                    \includegraphics[width = 0.45\textwidth]{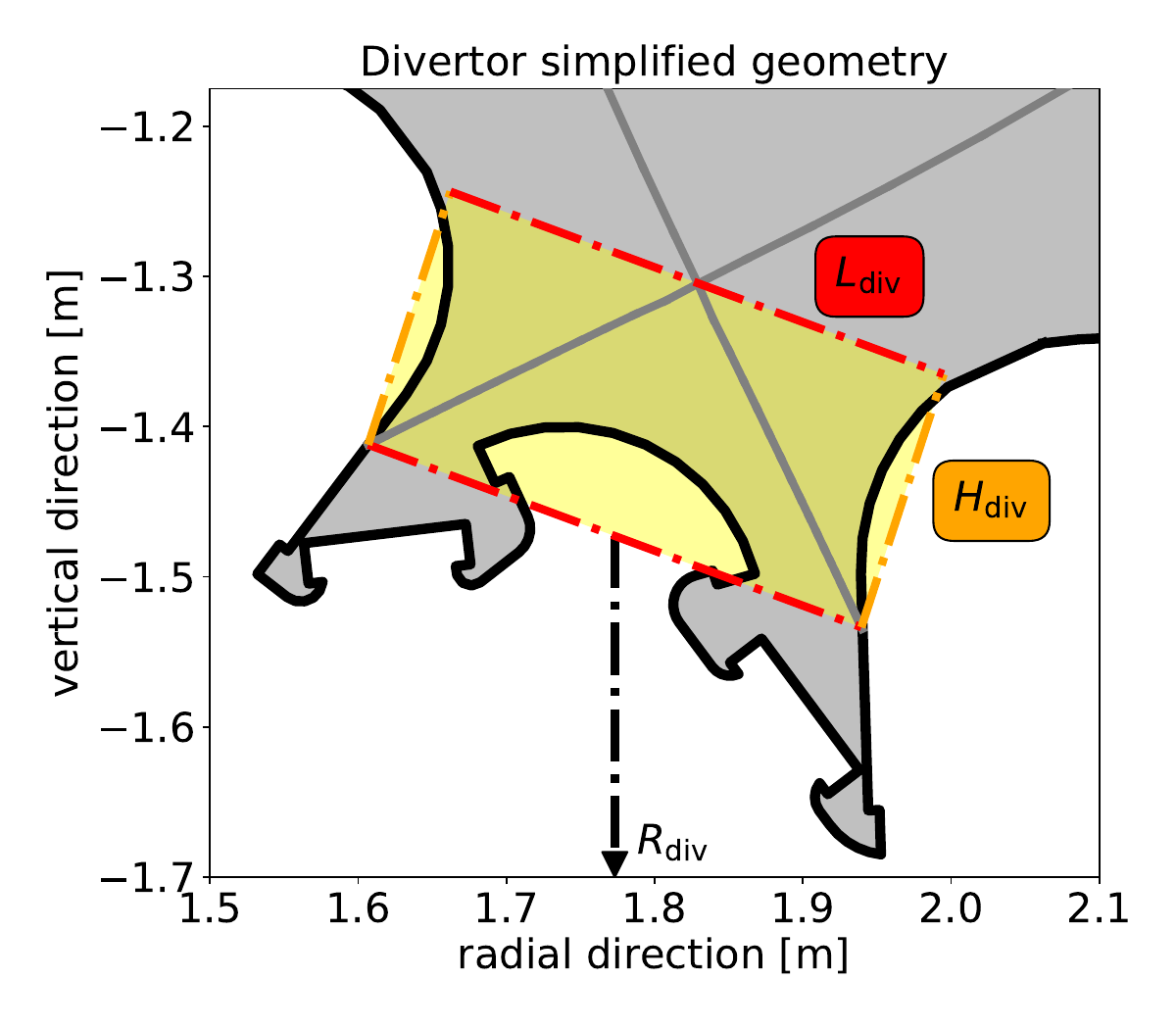}
                    \caption{View in the poloidal plane of the DTT's diverted single null plasma of Moscheni \textit{et al.} \cite{Moscheni_2022, Moscheni_2025}: wall in black; interior of the chamber in light grey; separatrix in dark grey; simplistic rectangular representation of the divertor in yellow; segment connecting inner, left, and outer, right, strike points in dash-dotted red (of length $\Ldiv$, with the radial location of its mid-point being $\Rdiv$); $\Hdiv$ is the distance of the X-point from it (orange).}
                    \label{fig:div_definitions}
                \end{figure}

                In equation (\ref{eq:VOFL_SOFL}), $\Ndiv$ is the number of active divertors and attains a value of 2 in the stellarators considered; in tokamaks \cite{Brunner_2018}: $\Ndiv = 1$ in single-null diverted plasmas, and disconnected double-nulls if featuring a power fall-off length significantly smaller than the separation of the two separatrices, i.e. $\lambda_q \ll \delta R_{\text{sep}}$; $\Ndiv = 2$ in double-nulls and disconnected double-nulls for which $\lambda_q \gtrsim \delta R_{\text{sep}}$. The toroidal coverage is estimated by $f_{\text{tor}}$: 50\% for W7-AS' and W7-X's island divertors, and 100\% otherwise. The remaining entries are defined via the yellow-shaded rectangle pictured in figure \ref{fig:div_definitions} (dash-dotted lines): $\Ldiv$ is the length of the segment connecting inner and outer strike points (red); $\Hdiv$ is the distance of the X-point from it (orange); $\Rdiv$ is the radius of the segment's mid-point (black). On average across the database\footnote{However, stellarators and, for instance, the WEST tokamak (section \ref{sec:results_Qpuff_volume}) violate these trends.} it holds that:
    
                \begin{equation}\label{eq:Ldiv_Hdiv_vs_a}
                    \begin{split}
                        \Ldiv & \sim 0.76 \times a \; ,\\
                        \Hdiv & \sim 0.35 \times a \; ,\\
                        \Rdiv & \sim 2.40 \times a \; ,
                    \end{split}
                \end{equation}

                and will be used in the following. The lack of closed field lines in linear plasma devices means that neither $\VCFL$ nor $\nesep$ are defined for such devices, and $\VOFL$ equals the chamber volume.
    
            \subsubsection{Physics quantities}\label{sec:methods_conventions_physics}

                Physics-based entries of the database (e.g. density) are left blank where not retrievable. In table \ref{tab:methods_database}, physics variables are:
                
                \begin{itemize}
                    
                    \item[\coloredsquare{white}] The number of L- and H-mode plasmas;
                    
                    \item[\coloredsquare{white}] The power crossing the separatrix $\Psep$;
                    
                    \item[\coloredsquare{white}] A generalised power fall-off length for tokamaks which combines Eich's regression \#14 \cite{Eich_2013}, i.e. $\lqEich [\text{mm}] = 0.63 \times (B_{\text{pol}}[\text{T}])^{-1.19}$, with the Scarabosio's correction factor $\fLH$ \cite{SCARABOSIO2013S426, Sieglin_2016, Reimerdes_2022, Faitsch_2018}, such that: $\fLH = 1$ for H-modes and $\fLH = 2$ for L-modes; not enough data are presently available to include the turbulence-driven broadening of \cite{Eich_2020} and/or Goldston's heuristic model \cite{Goldston_2012}, which are worth considering in the future nonetheless;
                    
                    \item[\coloredsquare{white}] The Greenwald density limit $\neGW [\permcube] = 10^{20} \times I_{\text{p}}[\text{MA}] / [\pi (a[\text{m}])^2]$ \cite{Greenwald_1988} and, additionally, the H-L density limit $\sim 0.45 \times \neGW$ from Eich \textit{et al.} \cite{Eich_2018} (HDL), valid for H-mode tokamak plasmas \cite{Bernert_2015}; stellarators instead abide to the Sudo density limit $\neSudo$ \cite{Sudo_1990};
                    
                    \item[\coloredsquare{white}] The separatrix electron density at the outer mid-plane $\nesep$;
                    
                    \item[\coloredsquare{white}] The opaqueness $\nesep \times a$, from a re-adaptation of the original definition from pedestal studies \cite{Mordijck_2020, REKSOATMODJO2021100971, Miller_2025}.
                
                \end{itemize}

            \subsubsection{Engineering actuators}\label{sec:methods_conventions_engineering}

                If not specified otherwise, the puffing rates of deuterium $\GammaD$ and seeded impurities $\GammaX$ are converted to their electron-equivalent $[\els]$ via the species atomic number \cite{Henderson_2023}.
                
                Only puffing-dominated cases are included in the database, where any core fuelling rate is negligible compared to $\GammaD$ \cite{Telesca_2022}, and other net particle sources/sinks (e.g. wall outgassing/retention \cite{Moulton2015Pumping}) can also be reasonably assumed as such.

                Table \ref{tab:methods_database} illustrates the vast range of cases included, with the actual rates spanning almost 5 orders of magnitude, and the ratio $\GammaD / \GammaX$ changing from impurity-dominated ($\ll 1$) to fuel-dominated ($\gg 1$ scenarios.
                
            \subsubsection{Detachment characterisation}\label{sec:methods_conventions_puffing_detachment}
                
                The data are extracted at one or more time instants during the detached phase of the outer target---the only target considered for reasons of data availability and, with the exception of spherical tokamaks \cite{Henderson_2025}, for usually being the most loaded \cite{Brunner_2018} and the last one to detach \cite{Krasheninnikov_Kukushkin_2017}. 
                
                Among the methods proposed \cite{Loarte_1998, Potzel_2014, Kallenbach_2015, Kallenbach_2016, Verhaegh_2023}, the degree of detachment (DOD) is \textit{qualitatively} referred to in the present work. Its original definition reads:

                \begin{equation}\label{eq:DOD}
                    \text{DOD} = \frac{\Gamma_{\text{D}^+}^{\text{2PM}}}{\Gamma_{\text{D}^+}^{\text{meas}}}
                \end{equation}

                where $\Gamma_{\text{D}^+}^{\text{2PM}} \sim \nesep^2$ is the integrated main ion flux at the divertor target as estimated from the 2-point model (2PM) \cite{stangeby2000plasma}---or from its readaptation to stellarators \cite{Feng_2011}, and $\Gamma_{\text{D}^+}^{\text{meas}}$ is the same flux actually measured (from experiments or simulations).
                
                Because only detached cases are here considered, DOD=1 specifically identifies the detachment onset during a density/puffing ramp, i.e. either (i) the appearance of ion-flux roll-over \cite{Potzel_2014, Fil_2020, Henderson_2023} or, if unavailable, (ii) the electron temperature dropping to $\sim 5$ eV around the outer strike point \cite{Leonard_2017, Chen_2018, Myatra_2023, HAVLICKOVA20151209, RUBINO2021100895}, or (iii) when explicitly indicated by the author. Instances more deeply detached than the onset are labelled by DOD$>$1.
                
                Main features of the DOD$>1$ and DOD=1 sub-sets are separately collated in table \ref{tab:methods_database}, and further details are given below.

        \subsection{Statistics}\label{sec:methods_statistics}

            The errors and limited data availability of a real-world database dictate the statistical methods required for an appropriate assessment---herein devised around the Occam's Razor heuristic \cite{occam}.

            \subsubsection{Statistical distinguishability}\label{sec:methods_statistics_distinguishability}
        
                Throughout the paper, the permutation test of difference in linear intercepts (with significance level of $5\%$) \cite{Fisher1992} is used to assess whether two sub-sets are statistically distinguishable---i.e. if there is supporting evidence in affirming that the two sub-sets differ with respect to a variable (or a collection of).

            \subsubsection{Physics-informed step-wise fitting method}\label{sec:methods_statistics_fitting}
    
                The suitability of the data fitting is indicatively quantified on the basis of $R^2$, though recognising that the model validation is the true judgement metric (section \ref{sec:methods_statistics_validation}).
                
                Generalising the recommendation of Zohm \textit{et al.} \cite{Zohm_2013} that fitting functions should be "motivated by experimental observations or physical arguments", but acknowledging the concerning finding of Body \textit{et al.} \cite{body2025} that "the apparent power law sensitively depends on the exact points used to build a scaling", a step-wise approach is herein adopted: (i) physics-centred correlations are first sought in ascending complexity; (ii) empirical calibration factors are added where improving validation; only in case (i) and (ii) were to fail, (iii) linear regression is resorted upon---preferring $N$ fitting parameters over $(N+1)$ to reduce the risk of over-fitting, modulo a marginal loss in $R^2$. And crucially, minimising the number of parameters maximises the data availability. Finally, (iv) well-established physical arguments are implemented in the laws discovered, to provide either upper bounds or meaningful simplifications.

            \subsubsection{Error sources}\label{sec:methods_statistics_errors}

                Disparate sources of error affect the data. From the perspective of the user, an unbiased, graphic data extraction error (digitisation error) of up to $\sim 15\%$ can be assumed affecting any datum pinpointed in a plot and not explicitly quoted in the text of the parent paper. Also unbiased, though not quantifiable, is any transcription/typographical error.
    
                Data-points from simulations feature convergence/discretisation errors---precisely defined \cite{Ghoos_2019, Boeyaert_2023}, but seldom assessed. Conversely, model errors are subtle, can change the detachment onset \cite{Paradela_Perez_2025} and can be noticeable even for a fixed scenario of a single machine \cite{Lore_2022}---but are constantly being researched with studies across \cite{Moscheni_2022, Moscheni_2025, Rivals_Soledge3x} and within \cite{Wu_2020, Makarov_2023, Wiesen_2025} simulation codes. The occasional struggle experienced in jointly replicating experimental data and puffing rates (or pressures) in present-day machines \cite{REIMOLD2015128, VAROUTIS201713, WIESEN2018174, Pan_2023} further testifies to the uncertainty of the puffing rates. 
    
                Regarding experimental data, Henderson \textit{et al.} \cite{HENDERSON2024101765} raise another issue specific to puffing: ``Requested flow rates [...] do not necessarily represent exactly the true injected flow rates''. Furthermore, diagnostic limitations affect the accuracy of experimental data---from magnetic reconstruction errors comparable to characteristic plasma length scales upstream (for $\nesep$) \cite{SILVAGNI2025101867, LaBombard_2011, Leonard_2017, Kallenbach_2018, Wigram2024Kinetic}, to Langmuir probes struggling in recombining detached plasmas downstream (for DOD=1 identification) \cite{MONK1997396, Ohno_2001, Okamoto_2006}. Unaccounted intrinsic impurities (e.g. carbon) further contribute to data scatter.

                To conclude with a positive note---the inherent uncertainty renders low-level operational differences indistinguishable, thereby allowing for the identification of robust scaling laws applicable across the entire database.

            \subsubsection{Non-comparability}\label{sec:methods_statistics_comparability}
    
                Common to both experimental and numerical data, an overarching confounding factor arises from the ``partiality'' of detachment (section \ref{sec:methods_conventions_puffing_detachment}). DOD varies continuously within $[1;+\infty)$, but quantifications of its value in the literature are practically inexistent. And instances at different, unquantified DOD lack direct comparability---points at different DOD would never precisely follow a scaling law which does not feature DOD among its variables.

            \subsubsection{Validation and definition of success}\label{sec:methods_statistics_validation}

                The sub-set of data at the detachment onset offers a solution to the above-mentioned complexities. Because DOD=1 by definition, such cases are directly comparable and, in principle, could be described by a scaling law agnostic of DOD. However, as for table \ref{tab:methods_database}, only a maximum of 45 DOD=1 data-points are available (reducing down to 16 in certain applications). \textit{Per se}, this is statistically unsuitable.
    
\begin{figure}
                    \centering
                    \includegraphics[width = 0.45\textwidth]{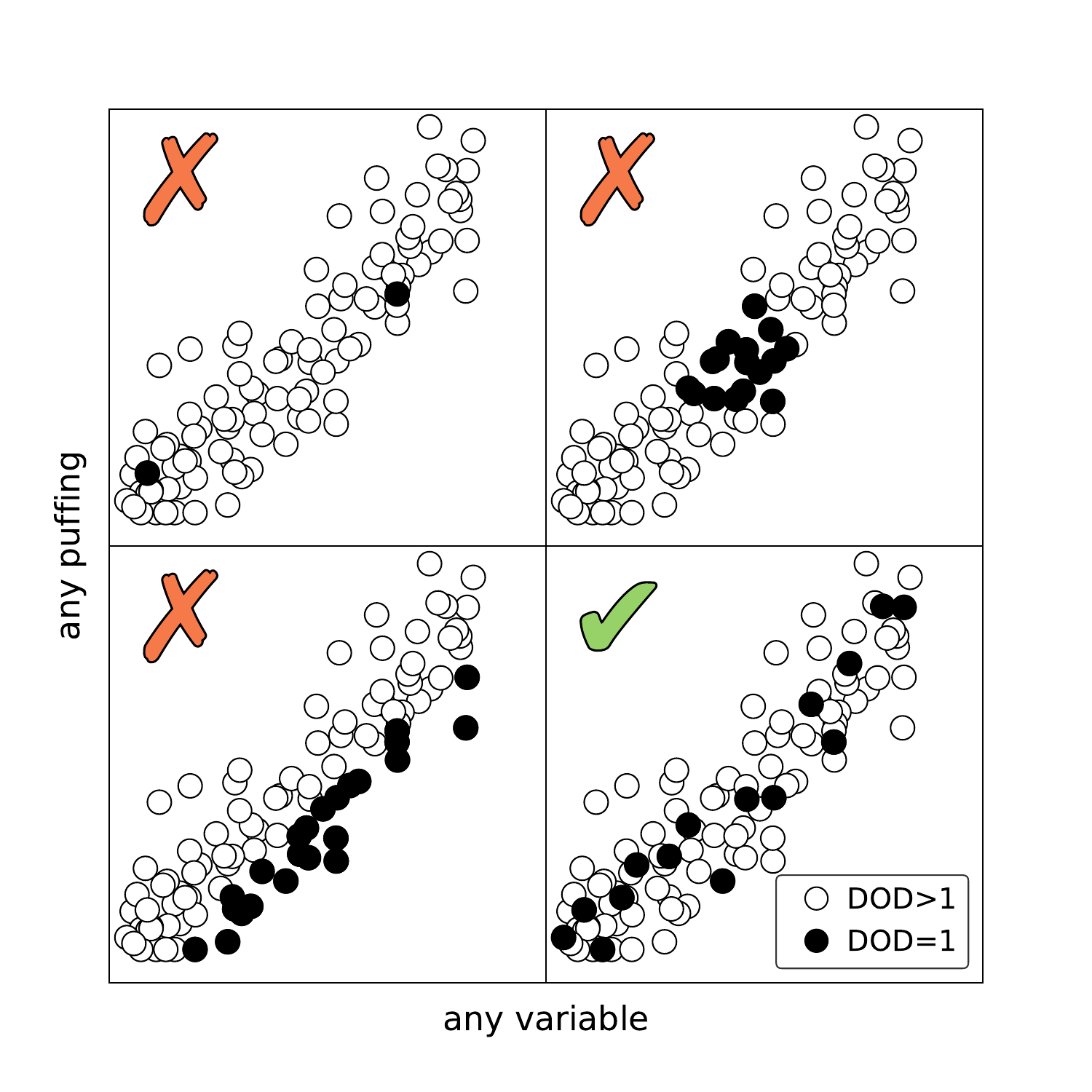}
                    \caption{Illustrative instances with dummy data where the strategy devised in section \ref{sec:methods_statistics_validation} does (green tick) and does not (red crosses) work. Top left: the number of DOD=1 points (black dots) is too small for validation. Top right: DOD=1 points are too clustered within the DOD$>$1 points in white. Bottom left: DOD=1 points are systematically below the rest. Bottom right: actual case, where DOD=1 points are well distributed (see text for details).}
                    \label{fig:scatter_cases}
                \end{figure}

                On the other hand, in the sense of section \ref{sec:methods_statistics_distinguishability}, it is anticipated that there is no evidence throughout this study that the sub-set of DOD=1 data lies systematically below its DOD$>$1 counterpart, (figure \ref{fig:scatter_cases}, bottom left), i.e. higher puffing does not necessarily imply higher DOD.
                
                A first clear example is provided by De Gianni \textit{et al.} \cite{DeGianni_2024}: even within the same device and for otherwise same input parameters, (i) $\GammaD = 3.6 \times 10^{22} \; \els$ leads to DOD=1 while (ii) the combination $\GammaD = 2.4 \times 10^{22} \; \els$ and $\GammaX = 0.3 \times 10^{22} \; \els$ results in DOD$>$1 despite the lower fuel, and total, puffing. The same is found by Lore \textit{et al.} \cite{Lore_2022}, with $\GammaD = 1.95 \times 10^{23} \; \els$ and $\GammaX = 0.40 \times 10^{23} \; \els$ resulting in a more deeply detached state than $\GammaD = 5.85 \times 10^{23} \; \els$ and $\GammaX = 0.06 \times 10^{23} \; \els$. Similar behaviours are even more likely to occur across scenarios and/or devices, when accounting for additional variables, non-linearities (e.g. see $\nesep$ in figure 4b of \cite{Lore_2022}) and due to errors (section \ref{sec:methods_statistics_errors})---each alone potentially turning the bottom-left situation in figure \ref{fig:scatter_cases} to the bottom-right one.
                
                Therefore, DOD$>$1 points are found to provide a source of \textit{unbiased} scatter around DOD=1 (figure \ref{fig:scatter_cases}, bottom right), presumably further aided by the scarcity of deeply detached scenarios (DOD$\gg$1). This allows for devising a robust strategy as follows.
                \\\\\textit{Database splitting}:
                \begin{itemize}
                    \item[\coloredsquare{white}] Training set: DOD$>$1 data are exclusively used to infer scaling laws;
                    \item[\coloredsquare{white}] Validation set: the scaling laws are validated against the unseen, comparable DOD=1 data.
                \end{itemize}
                The outcome of our study is therefore only applicable to DOD=1 data, and scored\footnote{The colour scoring is inspired by the finely-crafted table 2 of Oliveira and Body \textit{et al.} \cite{Oliveira_2022}.} based on criteria that reflect the impact of the sources of error of section \ref{sec:methods_statistics_errors}.
                \\\\\textit{Definition of success\footnote{For reference, a 2D edge plasma simulation would likely require $\lesssim 10\%$ of the total computational time to reconverge after adjusting a puffing rate by a factor 1.5.}}:
                \begin{itemize}
                    \item[\coloredsquare{MyGreenish}] for agreement within a factor 1.5;
                    \item[\coloredsquare{MyLightGreenYellow}] for agreement within a factor 3.0; 
                    \item[\coloredsquare{MyYellowOrange}] for agreement within a factor 6.0; \item[\coloredsquare{MyReddishOrange}] for anything worse. 
                \end{itemize}

                The average factor of agreement is quantified via the arithmetic and geometric averages\footnote{I.e. $\sum_{i=1}^nX_i / n$ and $(\prod_{i=1}^nX_i)^{1/n}$, respectively.}. The latter is a better representation in cases where a single instance of poor agreement significantly distorts the arithmetic average. 

                Finally, the above-mentioned examples from De Gianni \textit{et al.} \cite{DeGianni_2024} and Lore \textit{et al.} \cite{Lore_2022} also suggest that results from the present study only inform around a sufficient, though not necessary, condition for detachment access. Different fuel-impurity combinations leading to the same DOD might be possible.
    
        \section{Results}\label{sec:results}

            Plots in the following do contain both training (DOD$>$1) and validation (DOD=1) data-points for the sake of completeness. Data belonging to each family are highlighted in dedicated figure insets, for a visual demonstration of their resemblance to figure \ref{fig:scatter_cases} (bottom right). According to section \ref{sec:methods_statistics_fitting}, different scaling laws are built with incremental complexity and presented with their accompanying $R^2$.

            \subsection{Scaling laws for the fuel puffing rate}\label{sec:results_Qpuff}
    
                \subsubsection{Dominant role of geometry: correlation with volume}\label{sec:results_Qpuff_volume}
            
\begin{figure}
                    
                        \centering
                        \subfloat[]{\includegraphics[width = 0.45\textwidth]{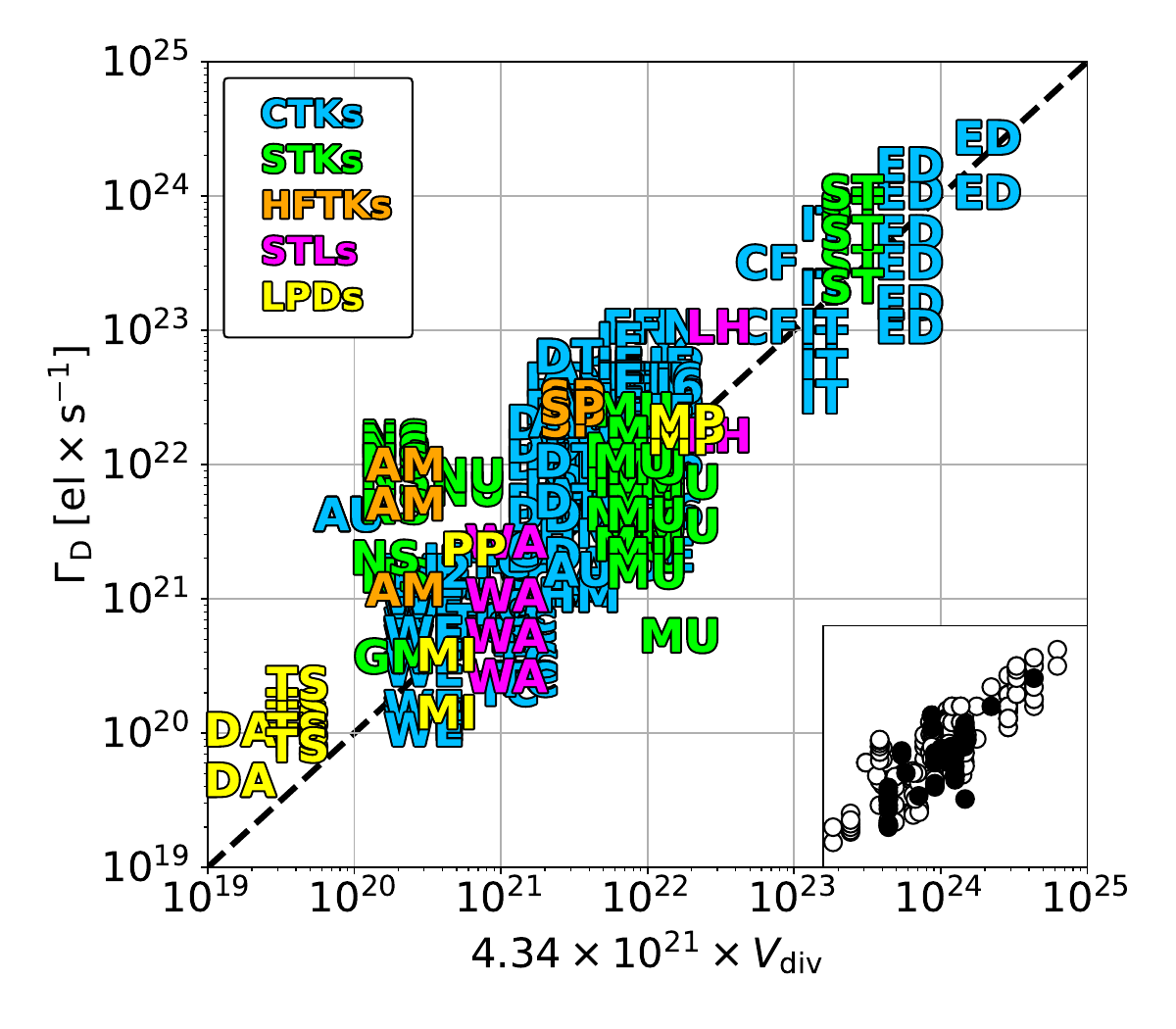}\label{fig:Vofl_vs_Qpuff}}
                        \\
                        \subfloat[]{\includegraphics[width = 0.45\textwidth]{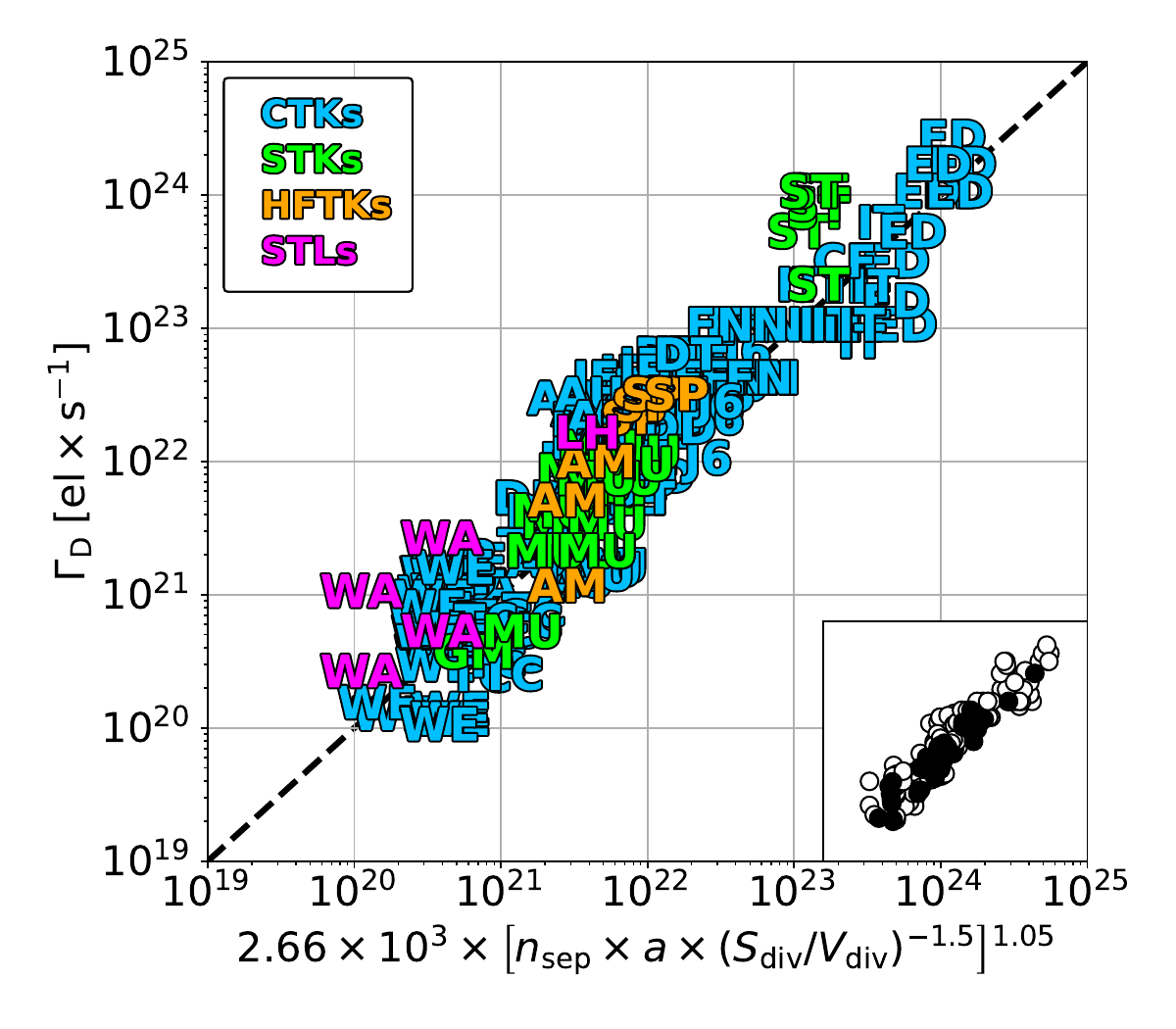}\label{fig:opacity_vs_Qpuff}}
                        
                        \caption{Results for the fuel puffing rate $\GammaD$ from: (a) the geometry-based equation (\ref{eq:Vofl_vs_Qpuff}) (VOL), exclusively depending on the divertor volume $\VOFL$; (b) equation (\ref{eq:V_OFL_S_OFL_1p5_times_opacity_vs_Qpuff}) (OPQ), where the opaqueness at the outer mid-plane separatrix $\nesep \times a$ appears alongside empirical calibration factors (exponents). Insets in the bottom right corners highlight points at DOD=1 (black) and DOD$>$1 (white) within the main plot. Scaling laws are derived from DOD$>$1 points only (training set).}
                        \label{fig:Qpuff}
                    \end{figure}

                    Figure \ref{fig:Vofl_vs_Qpuff} illustrates the strong linear correlation between the fuel puffing rate $\GammaD$ and the divertor volume $\VOFL$. Mathematically ($R^2 = 62\%$):
                
                    \begin{equation}\label{eq:Vofl_vs_Qpuff}
                        \GammaD^{\text{VOL}} [\els] = 4.34 \times 10^{21} \times \VOFL [\mcube] \; .
                    \end{equation}
                    
                    This relationship highlights the fundamental role geometry plays, in fact encompassing most of the machine-specific details and operational variations across conventional, spherical and high-field tokamaks, stellarators and even linear plasma devices (section \ref{sec:methods_database}). On the one hand, only retaining the conventional tokamaks---which dominate the dataset---does not produce a measurable variation (section \ref{sec:methods_statistics_distinguishability}), as the cumulative number of entries from the other devices is comparatively small. For the same reason, it is not currently possible to derive statistically significant laws specific to the remaining four machine types.

                    Of note---despite the renowned operational differences \cite{MATTHEWS2013S2}, no statistically significant discrepancy is found between carbon machines (e.g. JET and DIII-D \cite{Wang_2023}) compared to their metallic counterparts (e.g. the tungsten-walled AUG \cite{Kallenbach_2013}). The absence of a difference in this respect was similarly found by Eich \textit{et al.} \cite{Eich_2013} concerning the power fall-off length.
    
                    The linear plasma devices DT-ALPHA (``DA'' \cite{YOSHIMURA20251401015}) and TPD-Sheet IV (``TS'' \cite{Onda_2017}) lying above the trend is ascribed to under-estimating the total volume (the volumes of the interconnections between chambers are unaccounted for). Among the remaining notable outliers in figure \ref{fig:Vofl_vs_Qpuff}, both the family of metallic high-field tokamaks (i.e. Alcator C-Mod \cite{REKSOATMODJO2021100971} and SPARC \cite{Lore_2024}), the carbon-walled NSTX \cite{SOUKHANOVSKII2007432, Soukhanovskii_2009} and NSTX-U \cite{Chen_2018} are found to deviate from the average trend in the top-left area, MAST-U \cite{Moulton_2024} in the bottom-right one. Distinctive feature of such devices is their exceptional separatrix electron density $\nesep$---and corresponding exceptional opaqueness.

                \subsubsection{Secondary physics dependency: correlation with opaqueness}\label{sec:results_Qpuff_opacity}

                    Figure \ref{fig:opacity_vs_Qpuff} evidences how including $\nesep$ alongside geometry suffices in unifying the $\GammaD$ behaviour---in an instance where an empirically-calibrated physics-based dependency outperforms purely-empirical linear regressions (section \ref{sec:methods_statistics_fitting}). In particular, the opaqueness $\nesep \times a$ appears ($R^2 = 67\%$):
                
                    \begin{equation}\label{eq:V_OFL_S_OFL_1p5_times_opacity_vs_Qpuff}
                        \begin{gathered}
                            \GammaD^{\text{OPQ}} [\els] = 2.66 \times 10^3 \times \\
                            \times \left[ (\nesep [\permcube] \times a [\text{m}]) \times \left( \frac{\SOFL [\text{m}^{2}]}{\VOFL [\mcube]} \right)^{-1.5} \right]^{1.05} \; .
                        \end{gathered}
                    \end{equation}

                    Geometry-wise, equation (\ref{eq:V_OFL_S_OFL_1p5_times_opacity_vs_Qpuff}) relates compact divertors (high\footnote{For a fixed shape, e.g. a sphere of radius $r$, $S/V = 3/r$ and the ratio (and the compactness) increases as $r$ drops.} $\SOFL / \VOFL$) with a reduced $\GammaD$, for a given opaqueness. No dependence on aspect ratio is found, as expected from \cite{CHUANG2025101865}---where $A$ varies from 1.4 to 2.8, but opaqueness and fuelling do not.
    
                    Physically, that $\nesep$ matters across machines is natural. For instance, $\nesep \sim 8 \times 10^{19} \; \permcube$ in high-field tokamaks is a factor $\sim 3$ above that of H-mode plasmas in the other devices. Conversely, no such relationship of $\GammaD$ with the line-average electron density $n_{\text{avg}}$ holds. The puffed fuel is unable to penetrate the confined plasma deep enough for an effect on $n_{\text{avg}}$ to surface \cite{Mordijck_2020, REKSOATMODJO2021100971, Miller_2025, Leonard_2017}.
    
            \subsection{Scaling law for the impurity puffing rate}\label{sec:results_Zpuff}

                \subsubsection{Generalities}\label{sec:results_Zpuff_general}

\begin{figure}
            
                        \centering
                        \subfloat[]{\includegraphics[width = 0.45\textwidth]{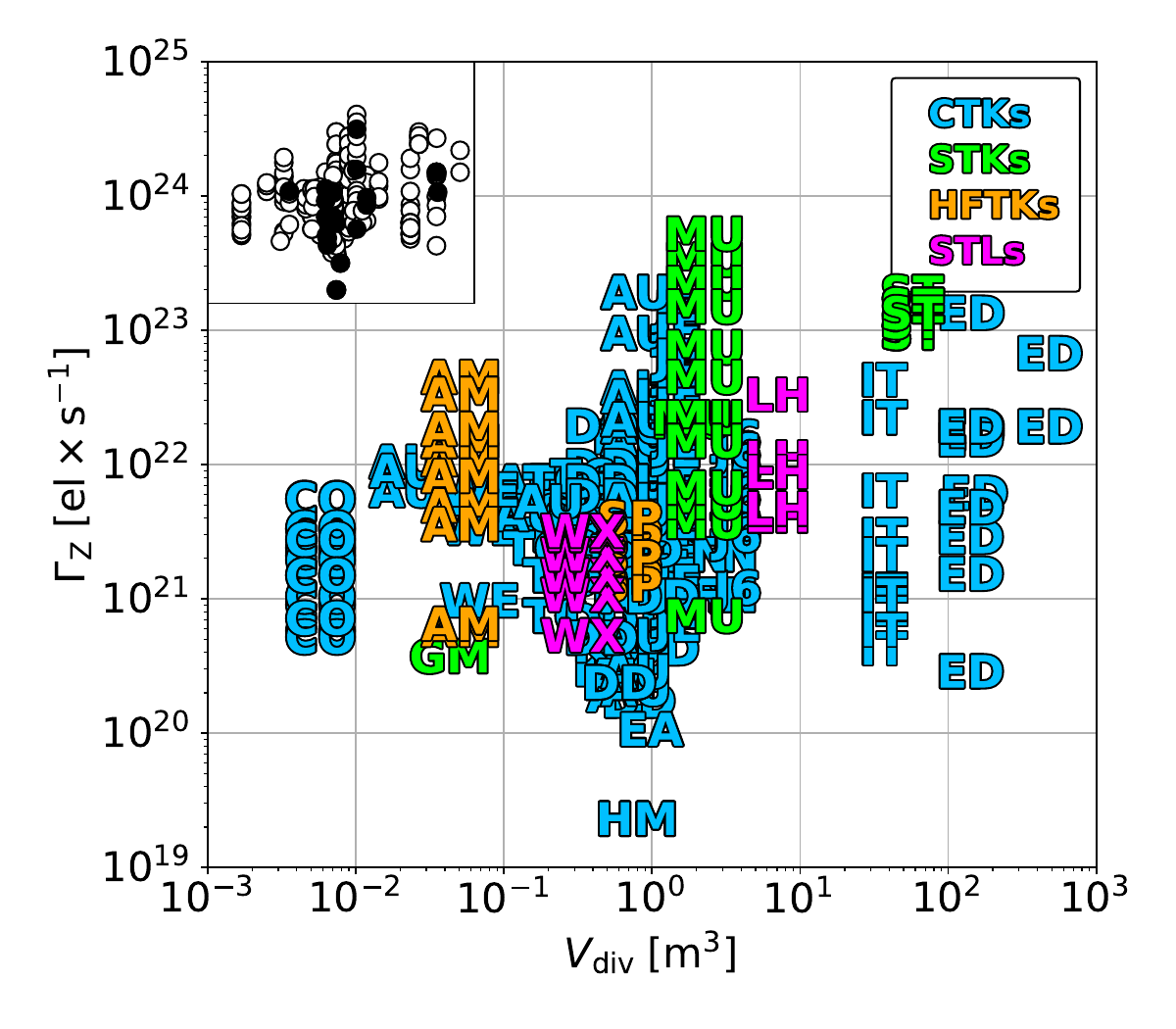}\label{fig:Vofl_vs_Xpuff}}
                        \\
                        \subfloat[]{\includegraphics[width = 0.45\textwidth]{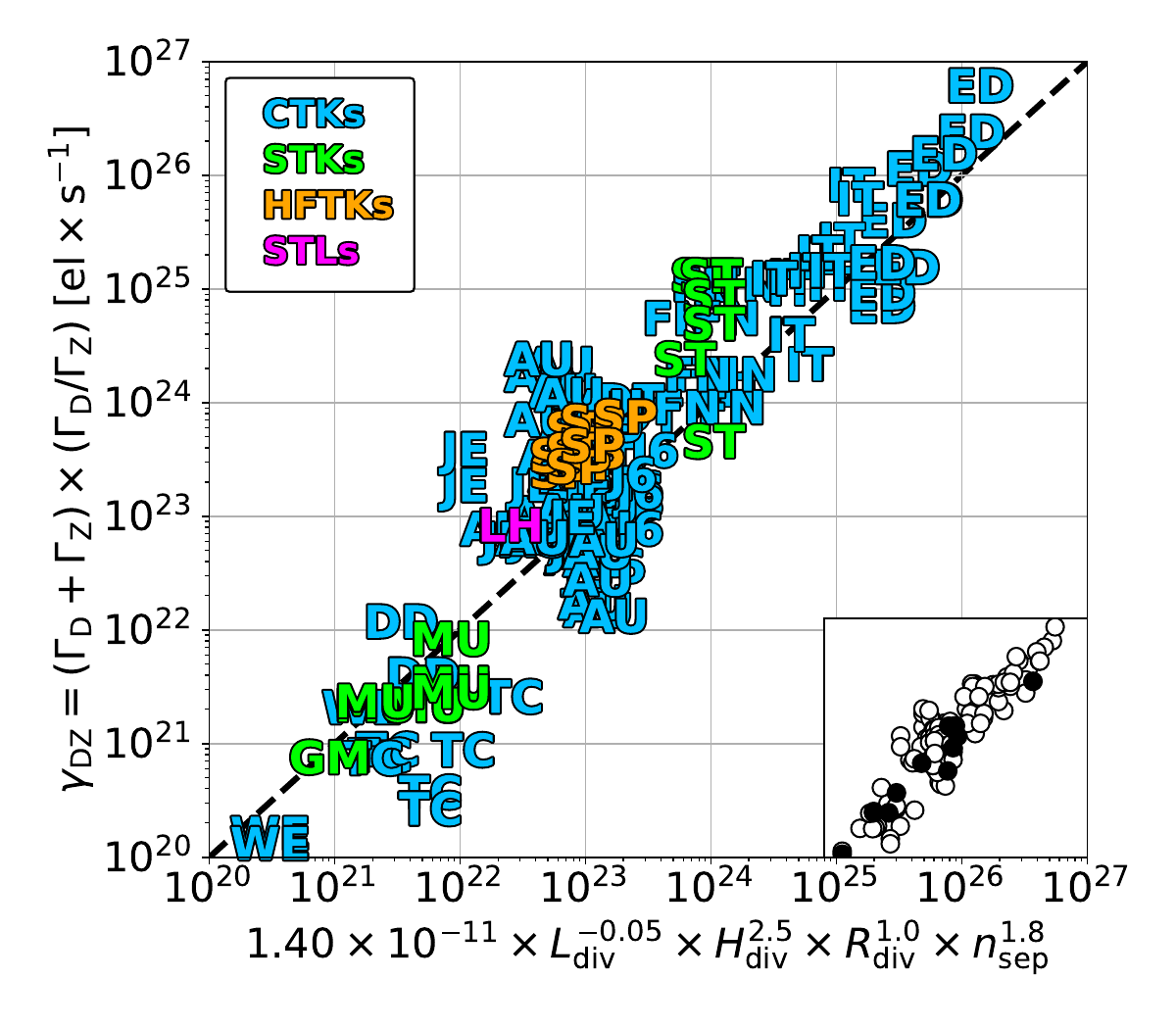}\label{fig:eq_gammaQZ_vs_MMM_composite}}
                        
                        \caption{(a) Lack of correlation of the impurity puffing rate $\GammaX$ with the divertor volume $\VOFL$. (b) Result of equation (\ref{eq:gammaDZ_aux}) for the non-linear function $\fQZ = (\GammaD + \GammaX) \times (\GammaD / \GammaX)$.} 
                        \label{fig:Xpuff}
                    \end{figure}

                    As shown in figure \ref{fig:Vofl_vs_Xpuff} and in contrast to $\GammaD$, the impurity puffing rate $\GammaX$ is not correlated to the divertor volume $\VOFL$. This feature could be explained, in part, by the residence time in the edge plasma $\tau \sim \Lconn / c_{\text{s}}$, where $\Lconn \sim \pi a B_{\text{tor}} / \Bpol$ is the connection length (outer mid-plane to target) and $c_{\text{s}} \sim T^{1/2}$ the plasma sound speed \cite{stangeby2000plasma}. Whether $T = \Tesep$ ($\sim R_0^{0.57}$ across the database) or $T \sim 5$ eV (in the detached zone) or a combination of the two is best suited to compute the sound speed, $c_{\text{s}}$ mildly increases with machine size at best. Therefore, $\tau$ must increase because of the leading dependence on connection length ($\Lconn \sim R_0^{0.64}$ in the database).
                    
                    The counter-intuitive, and somewhat surprising, result is that impurity puffing rates can be similar in a machine the size of COMPASS \cite{Komm_2021} ($R_0 \sim 0.6$ m, $\VCFL \sim 1 \; \mcube$ and $\Psep < 1$ MW) and in ITER \cite{VESELOVA2021100870} ($R_0 \sim 6$ m, $\VCFL \sim 900 \; \mcube$ and $\Psep \sim 100$ MW). The average seeding rates in the database indeed read $\GammaX \sim 2 \times 10^{21} \; \els$ and $3 \times 10^{21} \; \els$ for the two machines, respectively.
    
                \subsubsection{Linear regression of non-linear function}\label{sec:results_Zpuff_regression}

                    Operationally, the impurity seeding $\GammaX$ is usually added on top of fuel puffing $\GammaD$, and is hence expected to be a function of the $\GammaD$ level itself. Furthermore, the complexity of (i) the impurity transport \cite{Makarov_2023, Bufferand_2022}, (ii) the radiation efficiency \cite{Casali_2014, KALLENBACH2019166, Militello2022} and (iii) the involved interplay between $\GammaD + \GammaX$ and $\GammaD / \GammaX$ (figure 4b of \cite{Lore_2022}) hint at the presence of significant non-linearities---in the words of Body \textit{et al.} \cite{body2025}: "the impurity concentration required for detachment does not follow a simple power law". Last, (iv) it is noted that reactor-relevant puffing ratios $\GammaD / \GammaX \gg 1$ are achieved in bigger machines by means of high total puffing $\GammaD + \GammaX \gg 1$, and viceversa for smaller ones.

                    Therefore, a boolean-"AND" non-linear function able to account for all of the above is considered:

                    \begin{equation}\label{eq:gammaDZ_aux_definition}
                         \fQZ = (\GammaD + \GammaX) \times ({\GammaD} / {\GammaX}) \; ,
                    \end{equation}

                    which necessitates of the reactor-relevant condition $\GammaX > 0$. This allows for establishing the satisfactory linear regression pictured in figure \ref{fig:eq_gammaQZ_vs_MMM_composite} ($R^2 = 48 \%$):

                     \begin{equation}\label{eq:gammaDZ_aux}
                         \begin{gathered}
                             \fQZ [\els] = 1.40 \times 10^{-11} \times (L_{\text{div}} [\text{m}])^{-0.05} \times \\
                             \times (H_{\text{div}} [\text{m}])^{2.5} \times (R_{\text{div}} [\text{m}])^{1.0} \times (\nesep [\permcube])^{1.8} \; .
                         \end{gathered}
                    \end{equation}

                    Knowing the value of $\fQZ$, the impurity puffing rate is then computed by solving equation (\ref{eq:gammaDZ_aux_definition}) for $\GammaX$:

                    \begin{equation}\label{eq:GammaZ_final}
                        \GammaX [\els] = \frac{\left(\GammaD^{\text{OPQ}} [\els]\right)^2}{\fQZ [\els] - \GammaD^{\text{OPQ}} [\els]} \; ,
                    \end{equation}

                    where the fuel puffing rate is that of equation (\ref{eq:V_OFL_S_OFL_1p5_times_opacity_vs_Qpuff}). In contrast to the fuelling rate, but coherently with the aforementioned non-linearities, a non-trivial dependence of $\GammaX$ on $\nesep$ is found---both $\fQZ$ and $\GammaD$ in equation (\ref{eq:GammaZ_final}) depend on it.

                    Finally, differences in impurity species are also embedded in $\fQZ$---operation in smaller devices favours low-$Z$ impurities, and viceversa.

            \subsection{Simplified scaling laws}\label{sec:results_density_proxies}

                The presence of $\nesep$ in equation (\ref{eq:V_OFL_S_OFL_1p5_times_opacity_vs_Qpuff}) for $\GammaD$ and in equation (\ref{eq:GammaZ_final}) for $\GammaX$ introduces a layer of complexity \cite{Eich_2013, Kallenbach_2018} that simplified physics models \cite{body2025, Kallenbach_2016, Henderson_2025, Siccinio_2016, Kallenbach_2018, Cowley_2022, Silvagni2025Scaling, Goldston_2017}, AI-based ones \cite{DASBACH2023101396, Wiesen_2024} or other extrapolations \cite{SORBOM2015378_ARC, Scotti_2024} can synergistically address. Indeed, the value of $\nesep$ is usually quantified before puffing rates are \cite{SORBOM2015378_ARC, Rutherford_2024}.

                Nonetheless, physics-based approximations agnostic of $\nesep$ are sought in the light of section \ref{sec:methods_statistics_fitting}.

                \subsubsection{Tokamaks: density limits and power fall-off length}\label{sec:results_density_proxies_tokamaks}
                
                    An approximation for the tokamak's fuelling rate is found by rearranging equation (\ref{eq:V_OFL_S_OFL_1p5_times_opacity_vs_Qpuff}) by means of: (i) equation (\ref{eq:Ldiv_Hdiv_vs_a}); (ii) loosely capping $\nesep$ via HDL ($\lesssim 0.45 \times \neGW$, section \ref{sec:methods_conventions_physics}); (iii) the generalised power fall-off length (section \ref{sec:methods_conventions_physics}); (iv) the definition of poloidal magnetic field $\Bpol [\text{T}] = \mu_0 I_{\text{p}} [\text{A}] / (2 \pi a[\text{m}])$. Ultimately, the HDL loose upper bound reads:

                    \begin{equation}\label{eq:HDL_tokamaks}
                        \GammaD^{\text{HDL}} [\els] \sim 0.45^{1.05} \times \GammaD^{\text{GES}} [\els] \; ,
                    \end{equation}
                    
                    where $\GammaD^{\text{GES}}$ represents the Greenwald-Eich-Scarabosio (GES) strict upper bound ($\nesep < \neGW$):
                        
                    \begin{equation}\label{eq:V_OFL_S_OFL_1p5_times_opacity_vs_Qpuff_tokamaks}
                        \begin{gathered}
                            \GammaD^{\text{GES}} [\els] \sim 1.0 \times 10^{23} \times \\
                            \times (a [\text{m}])^{1.58} \times (\lqEich [\text{mm}] \times \fLH [-])^{-0.89} \; .
                        \end{gathered}
                    \end{equation}
    
                    It is worth noting that the HDL loose upper bound requires $\fLH = 1$, and that the GES strict upper bound only applies for DOD=1, at least in its present form. 
    
                    For $\GammaX$, the non-linearities involved in equation (\ref{eq:GammaZ_final}) mean that well-defined, general upper bounds can not be found. Therefore, $\nesep$ is directly proxied with the Greenwald density limit\footnote{Proxying via HDL would give a mere $\sim 20\%$ variation---negligible compared to the other intrinsic uncertainties and approximations involved in deriving the simplified laws.}, as accomplished in \cite{Eich_2013}. Leveraging the fact that $28 \times (\lqEich [\text{mm}] \times \fLH [-])^{-0.62} \gg 1$, equation (\ref{eq:GammaZ_final}) becomes:
    
                    \begin{equation}\label{eq:Zpuff_tokamaks_GW}
                        \begin{gathered}
                            \GammaX^{\text{GES}} [\els] \sim 3.6 \times 10^{21} \times \\
                            \times (a [\text{m}])^{1.51} \times (\lqEich [\text{mm}] \times \fLH [-])^{-0.27} \; ,
                        \end{gathered}
                    \end{equation}

                    applicable for both L- and H-modes.
                    
                    The negative correlations to $\lqEich$ further hint at the role of neutral penetration. Indeed, a figure of merit for neutral fuelling by means of puffing is the neutral mean free path (mfp) normalised to the minor radius---i.e. $\text{mfp}/a \sim \Bpol^{-1}$, close to the density limit \cite{Greenwald_2002}. Therefore, it holds that $\text{mfp}/a \sim \lqEich$, and the mean free path is approximately proportional to the power fall-off length. 

                \subsubsection{Stellarators: Sudo density limit}\label{sec:results_density_proxies_stellarators}

                    Algebra equivalent to that of section \ref{sec:results_density_proxies_tokamaks} with the Sudo density limit \cite{Sudo_1990} leads to the stellarator strict upper bound ($\nesep < \neSudo$):
                    
                    \begin{equation}\label{eq:V_OFL_S_OFL_1p5_times_opacity_vs_Qpuff_stellarators}
                        \begin{gathered}
                            \GammaD^{\text{Su}} [\els] \sim 2.2 \times 10^{22} \times \\
                            \times (a [\text{m}])^{1.58} \times \left( \frac{P_{\text{in}} [\text{MW}] \times B_{\text{tor}} [\text{T}]}{R_0 [\text{m}]} \right)^{0.53} \; ,
                        \end{gathered}
                    \end{equation}

                    and the proxy ($\nesep \sim \neSudo$):
    
                    \begin{equation}\label{eq:Zpuff_stellarators_SU}
                        \begin{gathered}
                            \GammaX^{\text{Su}} [\els] \sim 2.4 \times 10^{21} \times \\
                            \times (a [\text{m}])^{1.51} \times \left( \frac{P_{\text{in}} [\text{MW}] \times B_{\text{tor}} [\text{T}]}{R_0 [\text{m}]} \right)^{0.16} \; ,
                        \end{gathered}
                    \end{equation}
    
                    where $P_{\text{in}}$ represents the absorbed input power, $B_{\text{tor}}$ the toroidal magnetic field on axis. The reader is referred to section 2.4.2 of Greenwald \cite{Greenwald_2002} for uncertainties in stellarator density limits.

            \subsection{Validation}\label{sec:results_performance}

                The validation of the different scaling laws computed from the DOD$>$1 data-points is performed against the unseen DOD=1 sub-set. The results are summarised in table \ref{tab:validation} according to the indications of section \ref{sec:methods_statistics_validation}, with the asterisked entries being from experimental data and the superscripts specifying the L/H confinement mode (if available).
                
                No systematic bias is detected among the predicted values, as expected from section \ref{sec:methods_statistics_validation} (figure \ref{fig:scatter_cases}), and the validation is satisfactory as a whole. $\GammaD$ and $\GammaX$ predictions are found to be accurate within a factor $\sim 2$ on average (table \ref{tab:validation}, last row, with the geometric average within round brackets).

                Under-represented machine types are the linear plasma devices and the stellarators. For the former, only Magnum-PSI (\#31) suggests a potential for good agreement---but via the geometry-based equation (\ref{eq:Vofl_vs_Qpuff}) (VOL), the only one valid for linear plasma devices. No stellarator case at DOD=1 is available, and the applicability of the scaling laws is therefore questionable in their respect. However, it is noted that stellarators never represented an anomaly compared to the other devices within this study.

                \begin{table*}[htbp]
                    \centering
                    \renewcommand*\arraystretch{0.9}
                        \input{1-table-results}
                    \caption{Validation of the scaling laws computed with DOD$>$1 data against the unseen DOD=1 data (units $[\els]$, machine IDs from Appendix \ref{apx:nomenclature}). Colour scoring (section \ref{sec:methods_statistics_validation}): \coloredsquare{MyGreenish} for agreement within a factor 1.5; \coloredsquare{MyLightGreenYellow} within 3.0; \coloredsquare{MyYellowOrange} within 6.0; \coloredsquare{MyReddishOrange} anything worse; \coloredsquare{Silver} for the strict upper bound (equation (\ref{eq:V_OFL_S_OFL_1p5_times_opacity_vs_Qpuff_tokamaks})). White cells are either the actual puff rates, or predictions lacking the actual rate for comparison, or not available (--). Superscripts identify experimental data (*) and confinement mode (L)/(H). Last row: factor of agreement as arithmetic (geometric) average.}
                    \label{tab:validation}
                \end{table*}

                \subsubsection{Validation of scaling laws for fuel puffing rates}

                    Equation (\ref{eq:V_OFL_S_OFL_1p5_times_opacity_vs_Qpuff}) (OPQ) outperforms the geometry-based equation (\ref{eq:Vofl_vs_Qpuff}) (VOL) by virtue of the opaqueness dependence. This is not ubiquitous, but clearly present for the high-density tokamaks DTT (\#9), NSTX-U (\#32-33) and SPARC (\#34), where $\GammaD^{\text{VOL}}$ underestimates the true value by an order of magnitude. Equation (\ref{eq:Vofl_vs_Qpuff}) (VOL) tends to overestimate $\GammaD$ in MAST-U cases (\#21-30)---and equation (\ref{eq:V_OFL_S_OFL_1p5_times_opacity_vs_Qpuff}) (OPQ) addresses that shortcoming.
                    
                    The looseness of the HDL upper bound of equation (\ref{eq:HDL_tokamaks}) surfaces in 5/15 instances, where $\GammaD^{\text{HDL}}$ is overcome by the actual $\GammaD$. Although the limited data availability hinders further insight in such cases, estimates are presumably affected by, at least: (i) the very HDL definition in section \ref{sec:methods_conventions_physics} which involves the 0.45 factor ($\sim$ 0.4--0.5 in the original work \cite{Eich_2018}); (ii) the approximations adopted in the computation of $\GammaD^{\text{HDL}}$ (section \ref{sec:results_density_proxies_tokamaks}); (iii) the applicability of the HDL only to cases featuring a ballooning parameter $\sim$ 2.0--2.5 \cite{Eich_2018}, which is not herein quantifiable. That (i) matters is suggested by the JT-60SA entries (\#17 to \#20), for which $\nesep / \neGW$ grows from 0.20 to 0.34, hence approaching the $\sim 0.45$ mark and achieving increasingly better accuracy accordingly. On average, $\GammaD^{\text{HDL}}$ is found to be accurate within a factor 2.3 for H-modes. Crucially, it provides reliable estimates when $\nesep$ is not available and equation (\ref{eq:V_OFL_S_OFL_1p5_times_opacity_vs_Qpuff}) (OPQ) can not be used (e.g. \#23-24-28)---i.e. the very reason why the HDL equation (\ref{eq:HDL_tokamaks}) was devised in the first place.
                    
                    Finally, the Greenwald-Eich-Scarabosio strict upper bound (equation (\ref{eq:V_OFL_S_OFL_1p5_times_opacity_vs_Qpuff_tokamaks}), grey in table \ref{tab:validation}) sets a limit overcome neither by the actual $\GammaD$ nor by equation (\ref{eq:V_OFL_S_OFL_1p5_times_opacity_vs_Qpuff}) (OPQ) it is derived from.
    
                \subsubsection{Validation of scaling laws for impurity puffing rates}

                    The validation of impurity puffing rates is successful for conventional tokamaks despite the reduced data availability. Presumably due to tapering of the non-linearities and bearing intrinsic physical meaning (section \ref{sec:discussion_GES_essence}), the Greenwald-Eich-Scarabosio proxy in equation (\ref{eq:Zpuff_tokamaks_GW}) (GES) tends to even outperform its original version (equation (\ref{eq:GammaZ_final}), via $\fQZ$)---equally in H- and L-modes.
                    
                    Instead, impurity-seeded spherical tokamaks are under-represented. Mixed success is achieved in MAST-U cases, particularly at the extrema of DOD=1 with high seeding (\#21, $\GammaX = 1.7 \times 10^{23} \; \els$ with conventional divertor) and low seeding (\#30, $\GammaX = 7.0 \times 10^{20} \; \els$ with Super-X divertor). The fact that spherical tokamaks feature higher impurity puffing rates (figure \ref{fig:Vofl_vs_Xpuff}) might play a role, but attempts at including an aspect ratio dependence in equation (\ref{eq:gammaDZ_aux}) which validate well proved unsuccessful. However, the two instances of agreement (\#22-30) give confidence that spherical tokamaks could be described by simple scaling laws too.

        \section{Discussion}\label{sec:discussion}

            \subsection{Interpretation and applicability of the scaling laws}\label{sec:discussion_interpretation}

                Generally speaking, caution is required when interpreting any scaling laws, as correlation does not necessarily imply causation. In the present case, additional care is warranted because the laws are validated specifically for plasmas at detachment onset (DOD=1).

                For instance, a conceptual experiment in which $P_{\text{sep}}$ is increased would, in practice, cause the DOD=1 plasma to reattach (DOD$<$1). The comparable DOD=1 condition could only be restored by enhancing $\GammaD$ and/or $\GammaX$, which would in turn affect $\nesep$ and other correlated quantities \cite{Lore_2022}. As a consequence, disentangling the influence of individual, potentially co-varying parameters remains challenging. 
                
                This also highlights that our scaling laws describe transitions between DOD=1 states, rather than time evolution within a given discharge.
                
                In this sense, the present study should be viewed as a \emph{macroscopic} analysis---capturing trends \emph{across} devices, but not necessarily \emph{within} any single machine. The dominant dependencies in this regime are geometric---equation (\ref{eq:Vofl_vs_Qpuff}) being the leading example. As such, the scaling laws inform comparisons/extrapolations between machines, but do not necessarily describe the time-dependent response of a specific device unless, perhaps, its geometry changes significantly (see, e.g., the CRD in section \ref{sec:discussion_ADCs}).
                
\begin{figure}
                    \centering
                    \includegraphics[width = 0.45\textwidth]{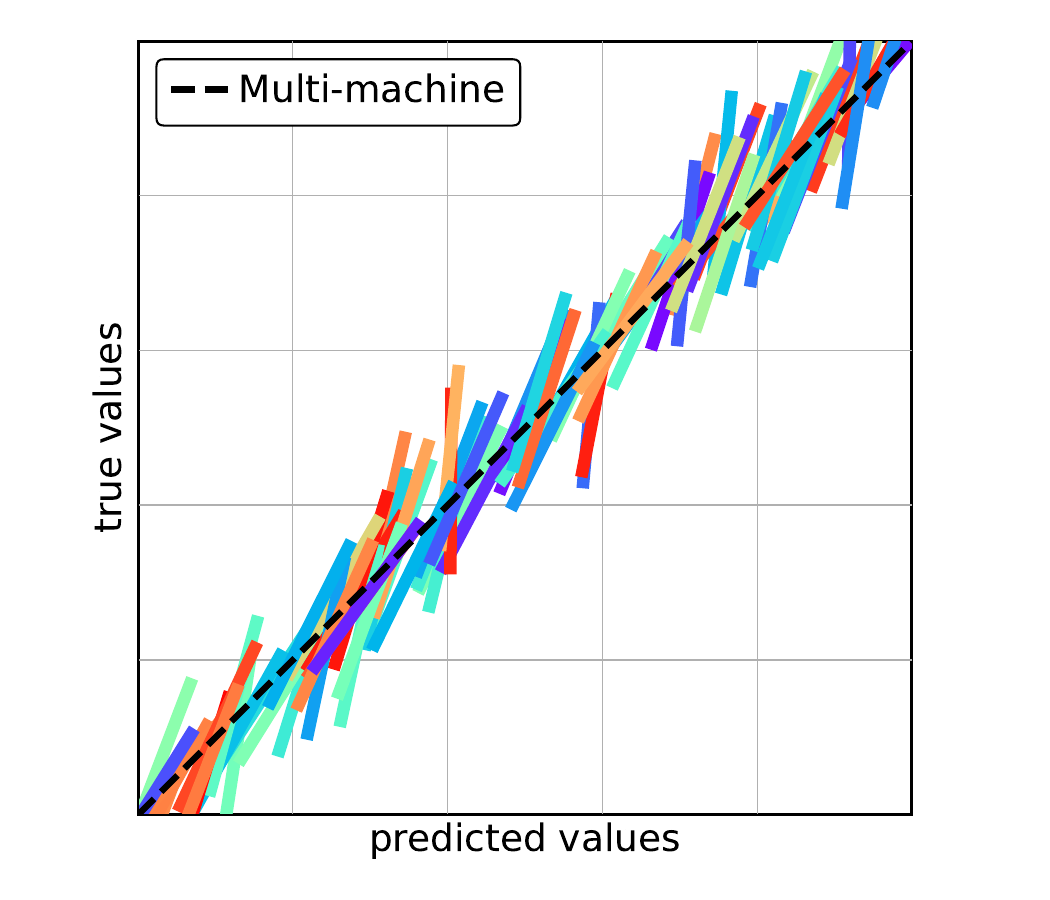}
                    \caption{Pictorial interpretation with dummy data of a prototypical scaling law---that is a macroscopic model of the trends across devices (a "barycentric average", dashed black line), but without necessarily being representative of each individual machine (coloured segments).}
                    \label{fig:Eich_segments}
                \end{figure}

                Figure \ref{fig:Eich_segments} pictorially illustrates this concept. Each coloured segment represents the behaviour of an individual machine, following its own local trend---simplistically assumed (log-)linear. The "barycentric average" of these segments gives rise to the global, multi-machine scaling law. Specialising the present framework to the \textit{microscopic} domain of a single device---where conditions vary continuously---forms part of our ongoing work, but is beyond the scope of the present one.

                Finally, it should be emphasised that the absence of a given parameter in the empirical laws does not imply that it plays no physical role. Rather, it indicates that its effect is not discernible. This can result from a combination of uncertainties (section \ref{sec:methods_statistics_errors}) and the aforementioned geometric dominance, which together tend to mask secondary dependencies.
                
                Further adding to the above, an often overlooked source of conditioning may arise from the \emph{operator bias}. Both experimental campaigns and edge plasma simulations naturally converge toward specific “recipes” of control parameters that ensure effective operation or stable detached performance---such as maintaining a target $n_{\text{sep}}$ or  numerical convergence. As a result, the accessible parameter space is not uniformly sampled. Examples include the $\GammaD / \GammaX$ ratios \cite{Lore_2022}, tailored impurity mixtures \cite{Henderson_2023}, and effective pumping speeds (section \ref{sec:pumping}).
                
                \textit{“Empirical universality” partly reflects the community’s convergence on similar operational strategies.}

            \subsection{Position with respect to existing knowledge}\label{sec:knowledge}

                Simplified analytical models have proven to be a reliable asset for 0.5D assessments of the edge plasma. The two fundamental approaches most commonly adopted \cite{Kallenbach_2018, Lengyel1981} are summarised by Body \textit{et al.} \cite{body2025}, and model inputs/outputs are exemplified by table 2 of Henderson \textit{et al.} \cite{Henderson_2025}.

                Concerning the latter \cite{Henderson_2025}, we note that retrieving $\GammaD$ and $\GammaX$ could be possible for tokamaks  (units $[\ats]$). Indeed, two approximations for as many variables are available.
                
                First, for single-null plasmas\footnote{Hence without the factor 2 at the denominator.} equation (8) of \cite{Henderson_2025} offers:

                \begin{equation}\label{eq:pdiv_kallenbach}
                    \pdiv = \frac{\GammaD + \GammaX + \Gamma_{\text{core}}}{\Seff} \times k_{\text{B}} T_{\text{pump}} \; ,
                \end{equation}

                where $\Seff \; [\mcube \times \text{s}^{-1}]$ is the effective pumping speed, $\Gamma_{\text{core}} \; [\ats]$ is the core fuelling rate (known and usually such that\footnote{E.g. $\Gamma_{\text{core}} \sim 10^{22} \; \ats$ corresponds to a mere 2--10\% of $\GammaD + \GammaX$ in ITER for scenarios at high-enough DOD \cite{Lore_2022, PITTS2019100696}; $\Gamma_{\text{core}} = 8 \times 10^{20} \; \ats$ is 4\% of $\GammaD + \GammaX$ in AUG \cite{SYTOVA201972, Senichenkov_2019, SENICHENKOV2023101361}; $\Gamma_{\text{core}} \sim 3 \times 10^{20} \; \ats$ (estimated from \cite{BARRETT2011789}) accounts for 6\% in MAST-U \cite{HENDERSON2024101765}; $\Gamma_{\text{core}} \sim 1 \times 10^{20} \; \ats$ (estimated from \cite{Coda_2017}) for $\lesssim 10$\% in TCV \cite{Yang_2023}. This also suggests that $\Gamma_{\text{core}}$ itself might follow a specific pattern.} $\Gamma_{\text{core}} \ll \GammaD + \GammaX$ for DOD$>$1), and $T_{\text{pump}} \; [\text{eV}]$ is the pump duct temperature (commonly $\sim 300 \;\text{K}$). These parameters are further commented in section \ref{sec:pumping}.
                
                Second, a relationship for the impurity concentration can be retrieved either (i) via equation (3) of \cite{HENDERSON2019147} for an estimate of (an upper bound of) impurity concentration in the divertor:
                
                \begin{equation}\label{eq:henderson_cz}
                    c_{\text{Z}} \lesssim \frac{\GammaX}{\GammaD + \GammaX} \; ,
                \end{equation}
                
                or (ii) via one among equations (3--5) of \cite{HENDERSON2021101000} (for AUG) or equation (6) of \cite{HENDERSON2021101000} (for JET), with then $\nesep \sim \pdiv^{0.31}$ \cite{Henderson_2025, Kallenbach_2018}. 
                
                While our 0D models utilise high-level, system-code-like parameters for input (see, e.g., table 3 of \cite{KOVARI20143054}), more sophisticated models like those of \cite{Henderson_2025, body2025} (0.5D) are designed to explore lower-level physics, and thus require more detailed inputs and underlying relationships. Given the proven usefulness of cross-comparisons between models of different fidelity \cite{Moscheni_2022, Moscheni_2025, HORSTEN2022101247} and dimensionality \cite{Myatra_2023}, one such exercise in the present context is believed to be valuable, but outside the scope of this work.
                
                For instance, our work uses upper bounds for $\nesep$ that are at times only loosely approached---a point of improvement for future studies. A cross-comparison with 0.5D models would involve testing their underlying scaling $\nesep \sim \pdiv^{0.31}$ \cite{Henderson_2025, Kallenbach_2018}, whose exponent is currently uncertain---values ranging from $\leq 0.0$ and up to $\sim 0.55$ have been found across different studies and seeding scenarios \cite{Kallenbach_2018, stangeby2000plasma, SCHWEINZER1999934, SILVAGNI2025101867, Silvagni2025Scaling}.
                
                The discrepancies between equation (\ref{eq:henderson_cz}) (i.e. equation (2) of \cite{HENDERSON2021101000}) and the trends in figure 4a of \cite{Lore_2022} are also noteworthy: $c_{\text{Z}} \sim \GammaX$ vs. $c_{\text{Z}} \sim \GammaX^{0.74}$ in the limit $\GammaX \ll \GammaD = \text{const}$; and $c_{\text{Z}} \sim \text{const}$ vs. $c_{\text{Z}} \sim \GammaX^{-0.24}$ if $\GammaD / \GammaX = \text{const}$, respectively.
                
                Because our equation (\ref{eq:gammaDZ_aux_definition}) for $\fQZ$ scales similarly to equation (\ref{eq:henderson_cz}), resolving these discrepancies (and those concerning $\nesep \sim \pdiv^{0.31}$) forms part of our current and forthcoming studies.

            \subsection{Puffing, pressure and pumping}\label{sec:pumping}

                Our results show that factor-2 accuracy can be obtained without explicit knowledge of $\Seff$, the effective pumping speed---a local quantity which would otherwise require a precise characterisation, depending on machine-specific hardware, its operating point, sub-divertor/duct geometry \cite{Varoutis_2024, Tantos_2024, TANTOS2025115021, KUKUSHKIN2007308, PSHENOV2025101851} and any wall contribution \cite{Moulton2015Pumping}. This suggests a degree of robustness in the proposed scaling laws.
                
                At the same time, $\Seff$ is recognised as a physically meaningful parameter whose role warrants further examination. 

\begin{figure}
                    \centering
                    \includegraphics[width = 0.45\textwidth]{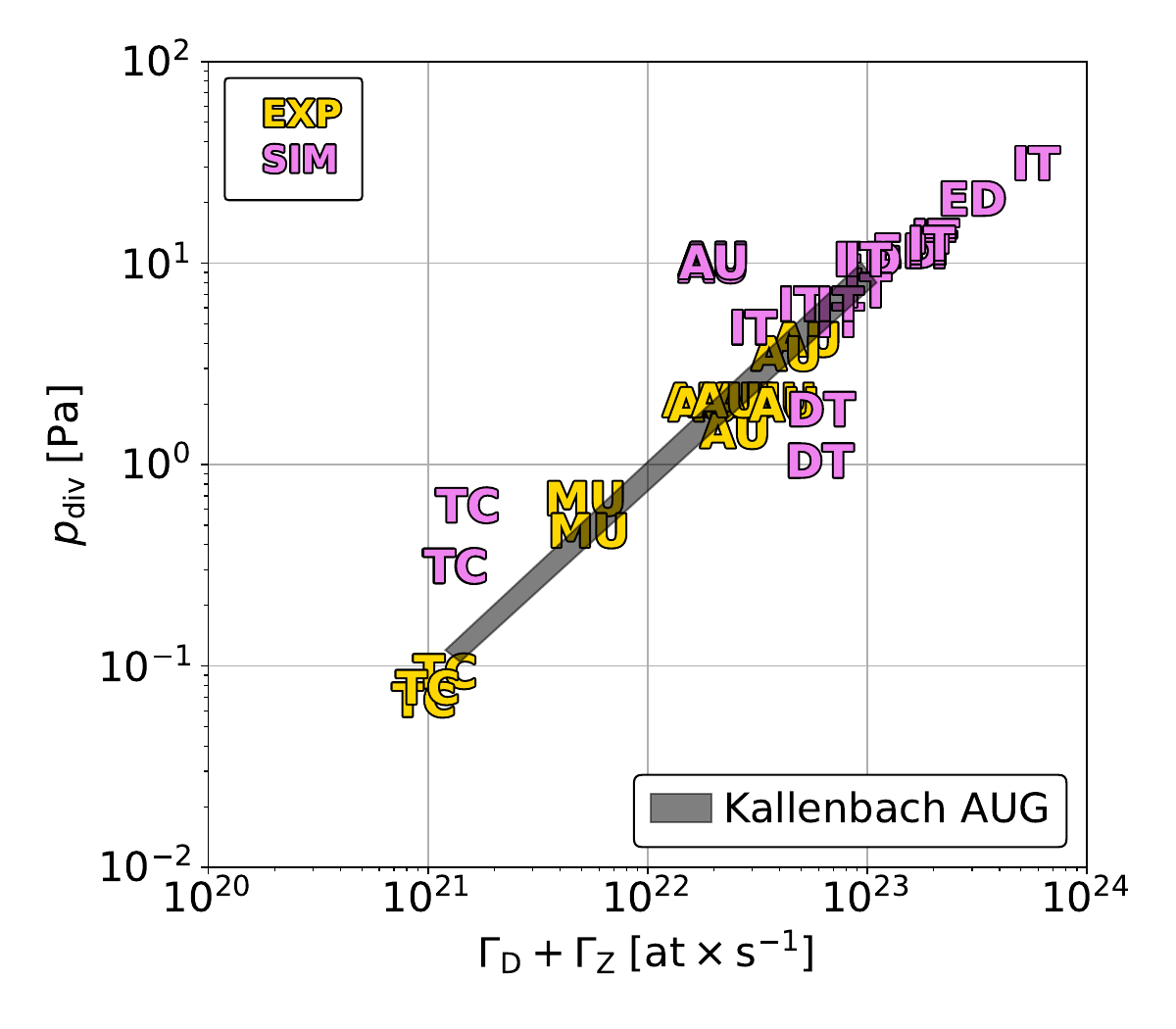}
                    \caption{Divertor neural pressure $\pdiv$ vs. total puffing rate for experimental (EXP) and simulated (SIM) cases in the present database. Overlaid in grey is the experimental result of Kallenbach \textit{et al.} \cite{Kallenbach_2018} in AUG (figure 5 therein), including both attached and detached cases.}
                    \label{fig:pdiv_scaling}
                \end{figure}

                To explore this aspect, figure \ref{fig:pdiv_scaling} plots $\pdiv$ against the total puffing rate (in $[\ats]$) across our database, overlaid with the experimental result of Kallenbach \textit{et al.} \cite{Kallenbach_2018} (figure 5 therein). The corresponding $\pdiv$-scaling reads ($R^2 = 47\%$):

                \begin{equation}\label{eq:pdiv_scaling}
                    \begin{gathered}
                        \pdiv [\text{Pa}] = 6.5 \times 10^{-23} \times \\
                        (\GammaD [\ats] + \GammaX [\ats]) \;, 
                    \end{gathered}
                \end{equation}

                and reproduces Kallenbach’s best-fit almost exactly---despite the additional presence of attached cases in \cite{Kallenbach_2018} (DOD$<$1).
                
                The sample size remains however limited due to the inclusion of $p_{\text{div}}$, resulting in not insignificant scatter. \textit{This strictly forbids any generalisation or claim of universality at this stage, as the observed pattern may in part be coincidental.}
                
                Still, the trend of figure \ref{fig:pdiv_scaling} suggests that the proportionality factor $\sim T_{\text{pump}} / \Seff$ in equation (\ref{eq:pdiv_kallenbach})---mainly depending on $\Seff$---is approximately constant within the uncertainties of the present study: $\Seff \sim 64 \; \mcube \times \text{s}^{-1}$ with $T_{\text{pump}} = 300$ K and neglecting $\Gamma_{\text{core}}$ (section \ref{sec:knowledge}). This might motivate the lack of a clear dependence on $\Seff$.
                
                However, it is remarkable that a similar value of $\Seff$---which factors in the contribution of the wall---is experimentally observed for detached cases in TCV \cite{Février_2020}, MAST-U \cite{HENDERSON2024101765}, and AUG\footnote{Some data might also be included in \cite{Kallenbach_2018}.} \cite{Kallenbach_2013, Reimold_2015, Kallenbach_2015, Henderson_2023}, and even for the attached scenarios of Kallenbach \textit{et al.} \cite{Kallenbach_2018}. Consistent behaviour is also found in simulations of EU-DEMO \cite{Subba_2021, Korzueva_2023} and ITER \cite{SYTOVA201972, SENICHENKOV2023101361, Lore_2022, SURESHKUMAR2024101780, Kaveeva_2020, VESELOVA2021100870}, which follow the same trend. Crucially, across the entire ITER database, a fixed pump capture coefficient of 0.72\% is adopted to explicitly reflect the realistic engineering constraint of $\Seff \sim 60 \; \mcube \times \text{s}^{-1}$ \cite{PITTS2019100696}.
                
                The only exceptions are \textit{simulated} cases in TCV \cite{Yang_2023}, AUG \cite{SENICHENKOV2023101361}, and DTT \cite{BALBINOT2021100952}, which deviate by up to a factor $\sim 5$ (DOD$>$1). Here a role might be played by both physical effects (e.g. plasma drifts not implemented or intrinsic code limitations \cite{Moscheni_2022, Moscheni_2025}) and geometric factors (e.g. the specific divertor location at which $p_{\text{div}}$ was sampled). Additional contributors likely include the \textit{estimated} wall recycling coefficients \cite{Yang_2023}, and/or the \textit{prescribed} pump capture coefficient. For instance, in AUG \cite{SENICHENKOV2023101361} the capture coefficient is not reported and thus cannot be assumed to correspond to the engineering $\Seff$. Instead, in DTT \cite{BALBINOT2021100952}, an 8\% capture coefficient was adopted---a value falling within the range identified by Tantos \textit{et al.} \cite{Tantos_2024} in the same machine, but for different underlying edge plasma simulations.
                
                The temperature $T_{\text{pump}}$ is also not systematically documented. It, however, influences $\pdiv$ according to equation (\ref{eq:pdiv_kallenbach}) and could vary with the pump duct location in the simulated domain \cite{Henderson_2025}---which does not usually include the entire sub-divertor volume \cite{KAVEEVA2021101030, PSHENOV2025101851}. This further complicates the interpretation, alongside the complexities discussed by Yang \textit{et al.} \cite{Yang_2023} in their equations (2--3).
                
                While numerical models have advanced considerably, capturing both pumping performance and plasma parameters simultaneously remains challenging \cite{Lore_2022}. These outliers are better interpreted as the consequence of differing modelling assumptions and boundary conditions.
                
                Overall, the mild variation of $\Seff$ across devices remains a striking feature---possibly reflecting the operator bias (section \ref{sec:discussion_interpretation}), whereby the community has converged on an effective pumping configuration that enables proficient operation within engineering and physical constraints.
                
                These observations are here reported as such---dedicated first-principle investigations \cite{TANTOS2025115021, Tantos_2024}, together with new experimental characterisations, will be needed to further comment on the matter.

            \subsection{Geometry}
            
            \subsubsection{Divertor volume over plasma volume}\label{sec:discussion_volumes}

                Although a growth of $\GammaD$ with the plasma volume $\VCFL$ would be intuitive, the present data suggest that the divertor volume $\VOFL$ actually dictates the correlation (equation (\ref{eq:Vofl_vs_Qpuff})). Indeed, the WEST (``WE'') tokamak is characterised by a peculiarly small $\VOFL/\VCFL$ ratio if compared to its peers (see, e.g., figure 1 in \cite{Yang_2024}). As a result, WEST simultaneously appears (i) consistent with other devices in figure \ref{fig:Vofl_vs_Qpuff} ($\VOFL \sim 0.1 \; \mcube$), and (ii) as a notable outlier\footnote{COMPASS also features a WEST-like $\VOFL/\VCFL$ ratio but no Q puff rate could be found in the literature.} when $\VCFL$ is considered instead (not shown)---in fact down to a factor 40 below the average trend. Therefore, either WEST is an anomaly within the entire family of magnetic confinement fusion devices, or $\VOFL$ is the leading metric. Evidence suggest the latter is the case, as divertor-specific geometrical parameters provide better correlations than plasma-specific ones throughout the study.

                Brida \textit{et al.} \cite{Brida_2025} demonstrate a correlation between the divertor radial extent (function of $\Ldiv$ and $\Hdiv$ herein) and the power fall-off length in the divertor private-flux region. This might explain a physical reason behind the above findings---transport from scrape-off layer to private-flux region.

            \subsubsection{Magnetic field geometry: the case of advanced divertor configurations}\label{sec:discussion_ADCs}

                Of appeal for future reactor designs, advanced divertor configurations (ADCs) \cite{MILITELLO2021100908, Lunt_PRL_CRD, Lee_Xpt_target_radiator_2025} optimise the poloidal/total flux expansion $f_{\text{X,R}}$, grazing angle $\theta_{\bot}$ and parallel connection length $\Lconn$ to ameliorate the power exhaust challenge.
                
                One such example is AUG's compact radiative divertor (CRD) of Lunt \textit{et al.} \cite{Lunt_PRL_CRD}, which approaches the geometrical limit $\lim_{a \rightarrow 0} (\SOFL / \VOFL) \sim \lim_{a \rightarrow 0} (8.3 / a) = +\infty$ (via equation (\ref{eq:Ldiv_Hdiv_vs_a})) and pushes equation (\ref{eq:V_OFL_S_OFL_1p5_times_opacity_vs_Qpuff}) to $\lim_{a \rightarrow 0} \GammaD = 0$. Although $\nesep$ data are not available in \cite{Lunt_PRL_CRD}, the CRD's $\GammaD \sim 0.4 \times 10^{22} \; \els$ sits at the lower end of the AUG's range in the present database ($\sim 2.0 \times 10^{22} \; \els$, on average) and, thus, is not inconsistent with the higher $\SOFL / \VOFL$ of the CRD. 
                
                Because the CRD's flux expansion approaches $+\infty$ in parallel to the divertor surface-to-volume ratio, $\SOFL / \VOFL$ can be thought as representing a proxy for the flux expansion. However, this does not apply at large, as the flux expansion can be magnetically tuned independently of $\SOFL / \VOFL$ (according to the definition in equation (\ref{eq:VOFL_SOFL})). All of the above affect the detachment onset---but it is not directly captured by the simplistic divertor approximation in figure \ref{fig:div_definitions}.

                Still, any effect mediated by $\nesep$ would be indirectly accounted for by the density dependences of equations (\ref{eq:V_OFL_S_OFL_1p5_times_opacity_vs_Qpuff}) and (\ref{eq:gammaDZ_aux}), at least partially. Let us consider the DOD=1 MAST-U cases from Moulton \textit{et al.} \cite{Moulton_2024} in table \ref{tab:validation}. The density drop from \#25 (conventional divertor, $\nesep = 6.0 \times 10^{18} \; \permcube$) to \#26 (Super-X ADC, $\nesep = 3.3 \times 10^{18} \; \permcube$) illustrates the facilitated detachment access via the Super-X ADC. The accompanying decrease in $\GammaD$ from $4.0 \times 10^{21} \; \els$ to $0.5 \times 10^{21} \; \els$ is partially captured by equation (\ref{eq:V_OFL_S_OFL_1p5_times_opacity_vs_Qpuff}) (OPQ) (from $2.2 \times 10^{21} \; \els$ to $1.4 \times 10^{21} \; \els$), but a difference persists---hence hinting at the role played by the unaccounted geometrical variables (e.g $\{f_{\text{R,X}} ;\: \theta_{\bot} ; \: \Lconn\}$). 
                
                Though not accomplished at this stage for reasons of data unavailability, explicitly including the above-mentioned variables is expected to enhance the predictive capability for ADCs. Similar arguments hold for variations of the size of edge islands in stellarators \cite{Feng_2016, PETERSON2025101868}.

            \subsubsection{Other geometrical features}

                Including the effect of further low-level geometric features---such as divertor closure \cite{cowley2024simulatingimpactbafflingdivertor}, puffing location \cite{Osawa_2024}, and vertical/horizontal target arrangement \cite{Loarte_2001}---into high-level scaling laws remains a challenging task.
                
                A useful proxy for divertor closure may be given by the ratio $\Ldiv/(2d_{\text{Xpt}})$, where $d_{\text{Xpt}}$ denotes the distance between the X-point and the closest wall element (not in the PFR). According to this metric: (i) in the limit $d_{\text{Xpt}} \rightarrow 0$, closure becomes absolute ($\infty$, i.e.\ the wall intersects the X-point); (ii) closure equals unity when $d_{\text{Xpt}} = L_{\text{div}} / 2$; and (iii) closure is $<1$ when $d_{\text{Xpt}} > \Ldiv/2$, corresponding to a very open divertor configuration (e.g. TCV \cite{FEVRIER2021100977, Reimerdes_2022}). Within this framework, $\GammaD$ would be expected to scale with $L_{\text{div}}/(2d_{\text{Xpt}})$ raised to a negative exponent (i.e. more closure implies smaller puffing to achieve DOD=1). Another approach for TCV is reported in \cite{Reimerdes2023, Tonello2025Modelling}.

                The influence of the divertor dome could, in principle, be captured by subtracting its volume from $\VOFL$, a correction that can be implemented when machine CAD data and/or edge plasma simulations are available.
                
                Regarding the puffing location, Osawa \textit{et al.} \cite{Osawa_2024} (figure 7, bottom left, in conjunction with their table 2) report a measurable variation of $\nesep$ with injection position---albeit within the range of the present uncertainties. Such dependence might explain the clustering of STEP (ST) data in figure \ref{fig:opacity_vs_Qpuff} of this work. However, a method to include varying puffing locations into multi-machine scaling laws is currently lacking.
                
                Similarly to the discussion in the previous section, these examples illustrate that while the inclusion of detailed geometric effects in global scalings is, in principle, feasible, it is not straightforward and will require further dedicated effort.

            \subsubsection{Limited plasma geometry}\label{sec:methods_limited}

                Plasma ramp-ups most commonly rely on an initial phase of limited plasma which can last several seconds \cite{PITTS2025101854}, and therefore require dedicated assessments because of the significant loads involved \cite{Kocan_2015}. Limited plasmas are not included in the current study as the leading dependences involve divertor parameters (section \ref{sec:discussion_volumes}). 

                It is however noted that, for similar \textit{plasma} volumes, $\GammaD$ puffing rates in detached limited plasmas can be within a factor 2 of diverted ones. TFTR \cite{bush1989neutral} compared to AUG \cite{Brida_2017, Lunt_PRL_CRD} and to WEST \cite{YANG2022101302, Yang2023Control}, and J-TEXT \cite{WU2023114023} compared to JFT-2M \cite{Kawashima_1999} and to TCV \cite{Harrison_2019} are two such examples.

                Including geometrical variables depending on the magnetic field (section \ref{sec:discussion_ADCs}) to avoid any divertor-related parameters would be worthwhile investigating for limited plasmas.

            \subsection{Exclusion of liquid lithium cases}\label{sec:methods_lithium}

                Despite the interest shown by the community \cite{deCastro_Li, Nallo_2022}, edge plasma simulations featuring liquid lithium plasma-facing components were discarded from the present study. We did not succeed in accounting for data from the insightful collection of works of Emdee \textit{et al.} \cite{Emdee_2021, Emdee_2023_1, Emdee_2023_2, Emdee_2024} for NSTX-U and of Islam \textit{et al.} \cite{ISLAM2022101292, Islam_2024} for FNSF---which invariably figured as outliers, despite having achieved detached conditions. 
                
                Such studies do cover, as a whole, a wide range of operational scenarios for a spherical and a conventional tokamak, respectively, and different liquid lithium models. Therefore, a possible cause of their unique behaviour could be the lithium-driven lower recycling---herein unaccounted for as a variable, but impacting on $\Seff$ (section \ref{sec:pumping}). Also potentially influential is lithium being an intrinsic impurity whose generation rate is dictated by the plasma, and not an externally-controlled extrinsic impurity.
                
                Therefore, only the scoping studies of Islam \textit{et al.} \cite{ISLAM2022101292, Islam_2024} not including lithium were here retained.

            \subsection{Lack of correlation with power}

                The divertor metrics which quantify the power exhaust challenge $\Psep / R_0$, $\Psep B_{\text{tor}} / R_0$ and $\Psep B_{\text{tor}} / R_0 / \nesep^2$ \cite{Rutherford_2024} do depend on plasma, rather than divertor, geometry. Additionally, $\nesep$ does not depend on heating power for a given $\GammaD$ \cite{Kallenbach_2018, Paradela_Perez_2025}. Therefore, the leading dependence on divertor, and not plasma, volume (section \ref{sec:discussion_volumes}) might be the reason why such $\Psep$-dependent divertor metrics fall short of correlating to the data better than, or even to the same degree of, equations (\ref{eq:V_OFL_S_OFL_1p5_times_opacity_vs_Qpuff}) and (\ref{eq:gammaDZ_aux}). The limited availability of $\Psep$ data might also be playing a role, and including $\Psep$ in the scaling laws would still be worthwhile pursuing.
                
            \subsection{Applicability across different scenarios}\label{sec:results_Qpuff_opacity_modes}
                
\begin{figure}
                    \centering
                    \subfloat[]{\includegraphics[width = 0.45\textwidth]{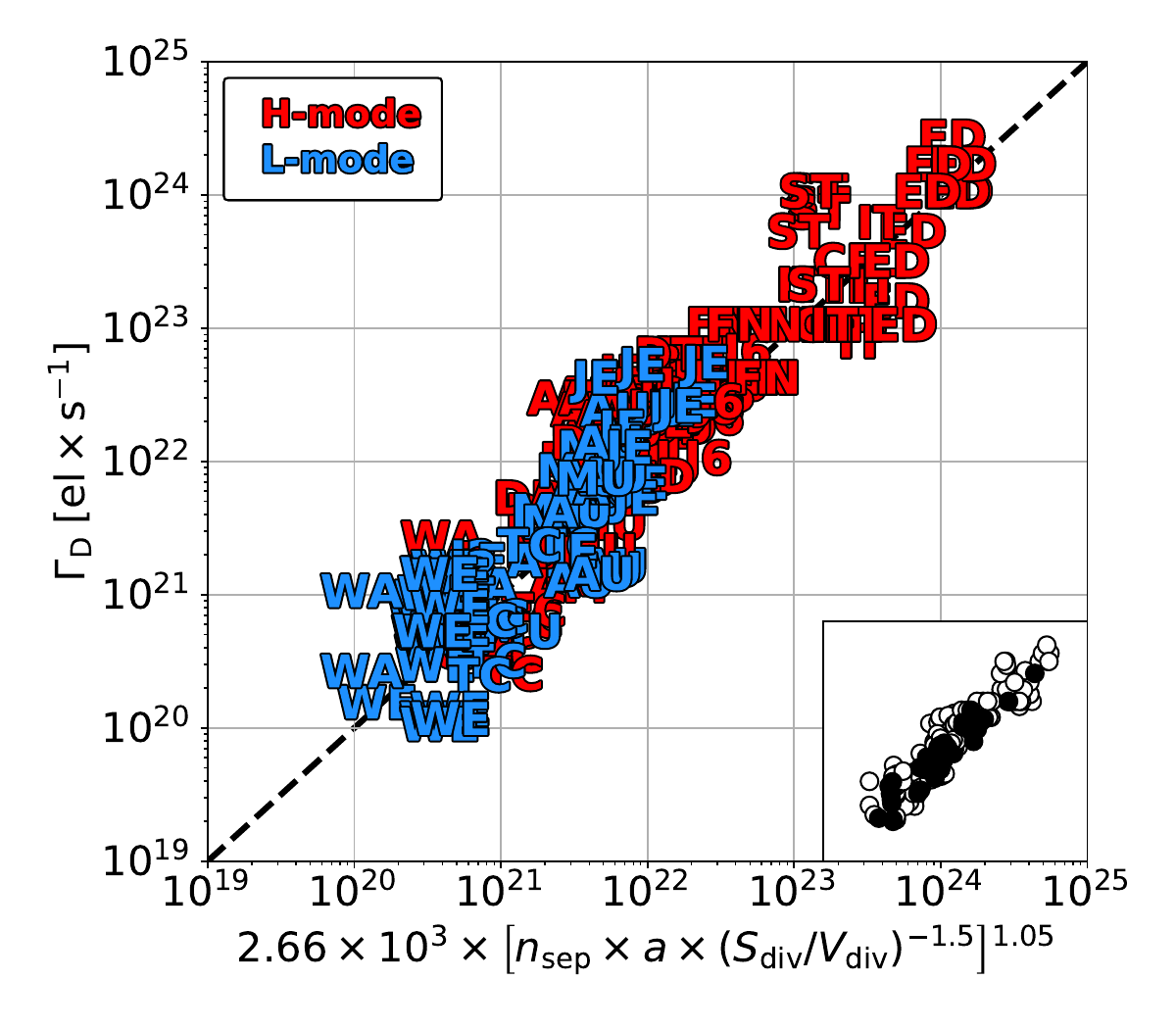}\label{fig:Qpuff_modes}}
                    \\
                    \subfloat[]{\includegraphics[width = 0.45\textwidth]{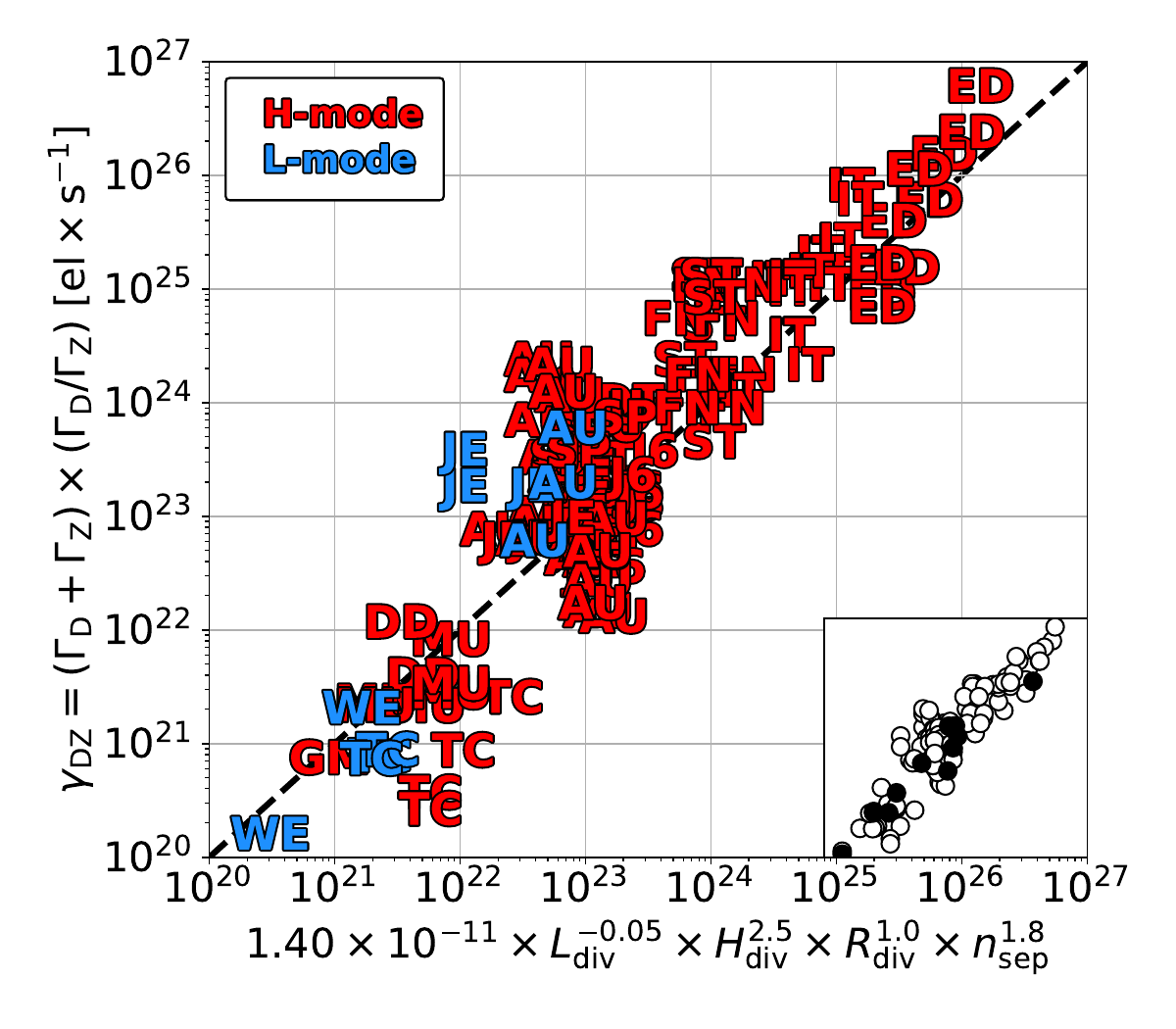}\label{fig:Zpuff_modes}}
                    
                    \caption{Analogous of figure \ref{fig:opacity_vs_Qpuff} for equation (\ref{eq:V_OFL_S_OFL_1p5_times_opacity_vs_Qpuff}), and of figure \ref{fig:eq_gammaQZ_vs_MMM_composite} for equation (\ref{eq:gammaDZ_aux}), but with the colouring which reflects the confinement regime.}
                    \label{fig:Qpuff_Zpuff_modes}
                \end{figure}

                The effect of a drastic change in opaqueness within a device is exemplified by the W7-AS stellarator (``WA'') in figure \ref{fig:opacity_vs_Qpuff} and \ref{fig:Qpuff_modes}. $\GammaD$ data from McCormick \textit{et al.} \cite{McCormick_2002_PRL} more than double from the average value of $0.6 \times 10^{21} \; \els$ in the low-density simil-L-mode regime (termed ``normal confinement'' \cite{McCormick_2002_PRL}, at abscissa $\sim 1 \times 10^{20}$), to $1.5 \times 10^{21} \; \els$ in the high-density simil-H-mode ones (``improved confinement'', at $\sim 7 \times 10^{20}$).
                
                Equations (\ref{eq:V_OFL_S_OFL_1p5_times_opacity_vs_Qpuff}) and (\ref{eq:gammaDZ_aux}) are also able to capture the smooth variation between L- and H-modes across all the other devices within figure \ref{fig:Qpuff_modes} and \ref{fig:Zpuff_modes}, respectively. This is quantitatively confirmed by the validation in table \ref{tab:validation}. For impurity puffing, a 20\% maximum deviation from H- to L-mode is expected, on paper, from equation (\ref{eq:Zpuff_tokamaks_GW}) and its factor $(\fLH)^{-0.27}$.

                Other concepts contemplated for future reactors \cite{SORBOM2015378_ARC, Rutherford_2024} are the I-mode \cite{Wagner_1982, Whyte_2010} and the negative triangularity \cite{Austin_2019}. On the one hand, $\GammaD$ data are not available in the I-modes of Reinke \textit{et al.} \cite{Reinke_2019}, and only one instance of negative triangularity exists in the database---that of Mombelli \textit{et al.} \cite{Mombelli_2025} (DOD$>$1). On the other hand, the L-mode-like particle confinement of the former \cite{Hubbard_2011}, and the L-mode-like pedestal absence of the latter \cite{Marinoni2021} suggest that the present results might also hold for I-modes and negative triangularity plasmas.

            \subsection{The essence of GES for impurity seeding}\label{sec:discussion_GES_essence}
              
\begin{figure}
                    \centering
                    \includegraphics[width = 0.45\textwidth]{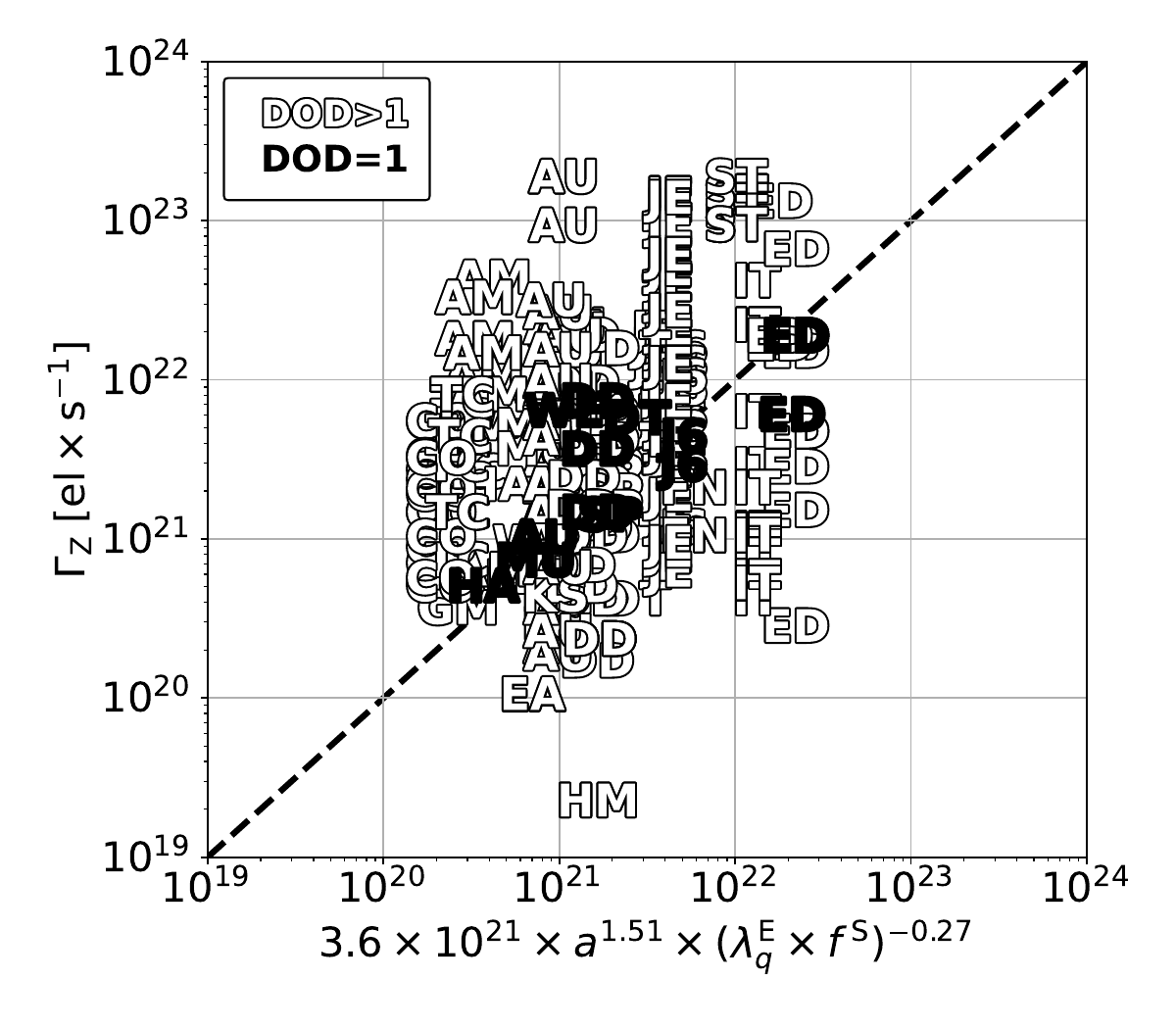}
                    \caption{Graphical representation of $\GammaX^{\text{GES}}$ in equation (\ref{eq:Zpuff_tokamaks_GW}), Greenwald-Eich-Scarabosio simplification of the scaling law for the impurity puffing rate. DOD=1 points (black) follow the scaling whereas the others do not, hence clarifying on the power of a physics-driven approach.}
                    \label{fig:GES_impurities}
                \end{figure}

                Figure \ref{fig:GES_impurities} pictorially illustrates the effect of the simplifications operated in section \ref{sec:results_density_proxies_tokamaks} to retrieve the Greenwald-Eich-Scarabosio impurity seeding rate $\GammaX^{\text{GES}}$: (i) in its present form, GES only applies for DOD=1 cases (black) while the reminder of the database (white, DOD$>$1) remains not sorted; (ii) although $\GammaX^{\text{GES}}$ follows a standard multiplicative functional form, (i) implies that attempts at finding such a correlation via linear regression are futile---and the approach based on $\fQZ$ is required (section \ref{sec:results_Zpuff_regression}); therefore, (iii) conversely to $\GammaD$ for which a scaling law could be directly found, $\GammaX$ is indissolubly tied to DOD.

                In conclusion, the physics-driven approach provided access to a region of the parameter space inaccessible to a purely-empirical regression.

            \subsection{Practical examples}\label{sec:results_examples}

                Table \ref{tab:examples} exemplifies the application of the scaling laws to conceptual power plants, hence predicting a combination of $\GammaD$ and $\GammaX$ sufficient to establish detachment (DOD=1)---with the exception of FSNF.
                
                DOD$>$1 cases for FNSF have already been thoroughly modelled. However, in the words of the author, Islam \textit{et al.} \cite{Islam_2024}: ``SOLPS is employed to identify the puffing and pumping levels [...]. Through an \textit{extensive set of scans}, it has been determined that a combination of Ne seeding ($1 \times 10^{20} \; \text{atoms s}^{-1}$) and $\text{D}_2$ puffing ($4 \times 10^{22} \; \text{atoms s}^{-1}$) meets all the FNSF demands [...]'' (italics added). In this instance (of, perhaps, DOD$\gtrsim$1) the values predicted for FNSF fall reasonably close to those found by the ``extensive set of scans'', and could have aided a multi-parameter scan which instead lasted a few months \cite{Islam2025_personal}. A reasonable match with a case of DOD$>$1 via scaling laws valid for DOD=1 is, however, not generalisable, as suggested by the violation of both HDL and GES upper bounds in table \ref{tab:examples} (last row).

                As for the other devices (yet to be thoroughly modelled), the stellarators Infinity Two \cite{Bader} and Stellaris \cite{LION2025114868} differ from the perspective of $\GammaD$ because of the expected $\nesep = 10^{20} \; \permcube$ for the former and $10^{19} \; \permcube$ for the latter. Although a Stellaris-like disparity from the output of equation (\ref{eq:Vofl_vs_Qpuff}) to that of equation (\ref{eq:V_OFL_S_OFL_1p5_times_opacity_vs_Qpuff}) is not unprecedented (e.g. \#17 in table \ref{tab:validation}), these results caution on the impact of $\nesep$ on the puffing rates and, ultimately, on the tritium fuel cycle \cite{Abdou_2021, Meschini_2023}.

                \begin{table*}[htbp]
                    \centering
                    \renewcommand*\arraystretch{0.9}
                        \input{2-table-examples}
                    \caption{Applications of the scaling laws to conceptual power plants in the shape of tokamaks and stellarators (asterisked).}
                    \label{tab:examples}
                \end{table*}

    \section{Conclusions}\label{sec:conclusion}

        A database gathering a representative collection of stably detached edge plasmas has been assembled, and the engineering actuators of fuel $\GammaD$ and impurity $\GammaX$ puffing rates investigated across magnetic confinement fusion devices. The inherent errors of experimental and numerical nature, alongside the confounding factor of the unquantified degree of detachment (DOD), force the devisal of a robust strategy---to build scaling laws on the sub-set of instances past the detachment onset (DOD$>$1) and validate against cases specifically at the onset itself (DOD=1)---the only truly comparable ones.

        While scoping DOD$>$1 data for scaling laws, the divertor volume is found to broadly determine the fuel puffing rate across all the machines, in fact encompassing most of the machine-specific details and operational variations---from desk-size linear plasma devices, through present-day stellarators and to future conventional and spherical tokamak pilot reactors.

        Refinement of this geometry-centred scaling law sees the appearance of the opaqueness at the outer mid-plane separatrix. With the addition of empirical calibration factors to an otherwise physics-based law, the inclusion of opaqueness unifies the behaviour by accounting for devices featuring an exceptional density (e.g. high-field tokamaks)---and that are accompanied by exceptional fuelling as a consequence.

        In regards to the impurity puffing rate, none of the above applies. $\GammaX$ is equally not correlated to geometry, opaqueness nor to any of the divertor metrics of the $\Psep/R_0$ family nor combination of. Motivated by the difference in the fuelling-seeding relationship between small and big devices, the non-linear variable $\fQZ = (\GammaD + \GammaX) \times (\GammaD / \GammaX)$ is defined. And a linear regression for $\fQZ$ does satisfactorily order the data-points, thus providing a route to the estimation of $\GammaX$, with $\GammaD$ being known on the ground of previous findings.

        The above suffices in fully characterising the engineering actuators both in L- and H-mode plasmas at DOD=1. However, the uncertainty involved in and the level of detail set by the separatrix electron density beg for simplification. By leveraging density limits, minimalistic laws as a function of machine design parameters---and of scalings for the power fall-off length at most---are established.

        Finally, the validation of the laws discovered is conducted on the DOD=1 set of data---and it is successful. Refining the $\GammaD$ scaling with the addition of opaqueness means that 43\% (13/30) of the available predicted values falls within a factor 1.5 of the actual value, 63\% (19/30) within a satisfactory factor 3, and the leftovers within a factor 6---while all respecting the Greenwald-Eich-Scarabosio strict upper bound $\GammaD^{\text{GES}}$. The H/L density limit is also shown to offer predictive capability for H-mode tokamak plasmas, accurate within a factor 2.3 and not depending on density.
        
        Scaling laws involving impurity seeding also successfully validate, modulo the reduced data availability. The simplified Greenwald-Eich-Scarabosio law $\GammaX^{\text{GES}}$ is found to outperform the original, non-linear version---in fact leading to 50\% of the predictions to match within a factor 1.5 (7/14), 29\% within a factor 3 (4/14).

        In conclusion, the present study demonstrates that a suitable combination of engineering actuators sufficient to access detachment can be accurately predicted for a given machine, on average within a factor 2. Moreover, simulations of future pilot plants are found to follow the extrapolations of present-day experimental data, hence adding on the confidence in the values predicted.
        
        The laws determined find immediate applicability in system codes and high-level scoping studies, alongside providing guidance in low-level edge plasma modelling---to ultimately enable the rapid conceptual design and sizing of an appropriate tritium fuel cycle architecture and particle exhaust systems.

        What these laws do \emph{not} guarantee is an accurate description of the microscopic behaviour within a given device. Instead, they offer a macroscopic view across multiple machines, capturing system-level trends rather than machine-specific dynamics.

        In perspective, contributions to expand/refine/revise the database (available at \cite{zenodo_repo}) are \textit{strongly encouraged}. This would enable a better characterisation of the categories under-represented in the validation, which would be needed to assess the applicability of the laws to spherical tokamaks (impurity-wise and at the inner target), linear plasma devices, and advanced divertor configurations in general. Because the corresponding DOD$>$1 data tend to follow the others, an extension of our scaling laws should apply for these cases. For the excluded liquid lithium cases an \textit{ad hoc} approach would instead need development.

        Though intriguing, further speculations about the relationship between divertor neutral pressure, puffing, and pumping across devices is not possible at this stage. Nevertheless, this interplay---together with the potential influence of the operator bias---remains an important subject for future dedicated work.
        
        Our forthcoming studies focus on designing new laws descriptive of the behaviour of the individual machine. Additionally, in parallel to investigating the physical reasons behind the laws uncovered, revisiting the present results by accounting for the degree of detachment would give the opportunity to predict puffing rates for scenarios more deeply detached (DOD$>$1)---near reactor-relevancy. Whether including a quantification of ``complete'' detachment and/or MARFEs would allow to set an absolute inviolable limit to puffing rates remains to be seen.

    \section{Acknowledgements}\label{sec:acknowledgements}

        This work has been (part-) funded by the EPSRC Energy Programme [grant number EP/W006839/1].
        
        The authors would like to acknowledge the diligence of their colleagues across the scientific community for meticulously reporting valuable data in their publications. 
        
        J.D. Lore, E. Tinacba, O. Myatra, K. Verhaegh, M.S. Islam, J. Brownfield, M. Wigram and A. Kallenbach are also warmly thanked for providing useful pieces of information during the development of this work, and so are the teams of ASDEX Upgrade and UKAEA for the constructive feedbacks during seminars.
        
        Stimulating discussions with T. Eich and D. Silvagni that helped interpreting the results deserve being emphatically acknowledged.
        
        Last but not least, a special thank goes to S. Cirone (RUP) and C. Soika (Comm. Rex) for the ever-present support.


    \appendix

    \section*{Appendix}

    \section{Nomenclature and sources}\label{apx:nomenclature}

        Table \ref{tab:references} reports the machines figuring in the study, their unique ID and the scientific papers where data have been extracted from.

        \begin{table*}[htbp]
            \centering
            \renewcommand*\arraystretch{0.9}
                \input{3-table-appendix}
            \caption{Machine, corresponding ID and references used in the work.}
            \label{tab:references}
        \end{table*}
    
    \bibliographystyle{IEEEtran}
    \bibliography{processed_references.bib}

\end{document}

%% file: 1-table-results.tex
\begin{tabular}{cccccccccc}
\toprule
\multirow{2}{*}{\#} & \multirow{2}{*}{ID [ref.]} & \multirow{2}{*}{$\Gamma_{\text{D}}$} & Eq. (\ref{eq:Vofl_vs_Qpuff}) & Eq. (\ref{eq:V_OFL_S_OFL_1p5_times_opacity_vs_Qpuff}) & Eq. (\ref{eq:HDL_tokamaks}) & Eq. (\ref{eq:V_OFL_S_OFL_1p5_times_opacity_vs_Qpuff_tokamaks}) & \multirow{2}{*}{$\Gamma_{\text{Z}}$} & Eq. (\ref{eq:GammaZ_final}) & Eq. (\ref{eq:Zpuff_tokamaks_GW}) \\
 & & & VOL & OPQ & HDL & GES & & via $\fQZ$ & GES \\\midrule
1 & AU$^{\text{(L)}}$* \hfill \cite{Brida_2017} & 7.0E+21 & \cellcolor{MyLightGreenYellow}3.4E+21 & \cellcolor{MyLightGreenYellow}4.0E+21 & \cellcolor{white}-- & \cellcolor{Silver}8.1E+21 & 9.8E+20 & \cellcolor{MyLightGreenYellow}4.6E+20  & \cellcolor{MyGreenish}8.2E+20 \\
2 & AU$^{\text{(L)}}$* \hfill \cite{Potzel_2014} & 1.2E+21 & \cellcolor{MyLightGreenYellow}3.4E+21 & \cellcolor{MyLightGreenYellow}3.4E+21 & \cellcolor{white}-- & \cellcolor{Silver}1.0E+22 & -- & \cellcolor{white}4.4E+20  & \cellcolor{white}8.8E+20 \\
3 & AU$^{\text{(L)}}$ \hfill \cite{WU2023114023} & 4.0E+21 & \cellcolor{MyGreenish}3.4E+21 & \cellcolor{MyGreenish}3.3E+21 & \cellcolor{white}-- & \cellcolor{Silver}1.0E+22 & -- & \cellcolor{white}4.4E+20  & \cellcolor{white}8.8E+20 \\
4 & AU$^{\text{(L)}}$ \hfill \cite{Wu_2021} & 5.8E+21 & \cellcolor{MyLightGreenYellow}3.4E+21 & \cellcolor{MyGreenish}5.3E+21 & \cellcolor{white}-- & \cellcolor{Silver}1.0E+22 & -- & \cellcolor{white}4.8E+20  & \cellcolor{white}8.8E+20 \\
5 & CF$^{\text{(H)}}$ \hfill \cite{MAO20151233} & 1.0E+23 & \cellcolor{MyLightGreenYellow}6.6E+22 & \cellcolor{MyLightGreenYellow}1.6E+23 & \cellcolor{MyLightGreenYellow}1.7E+23 & \cellcolor{Silver}4.0E+23 & -- & \cellcolor{white}6.5E+21  & \cellcolor{white}8.9E+21 \\
6 & DD$^{\text{(H)}}$* \hfill \cite{Wang_2023} & -- & \cellcolor{white}2.0E+21 & \cellcolor{white}3.0E+21 & \cellcolor{white}1.3E+22 & \cellcolor{Silver}2.9E+22 & 1.4E+21 & \cellcolor{MyLightGreenYellow}2.2E+21  & \cellcolor{MyGreenish}1.6E+21 \\
7 & DD$^{\text{(H)}}$* \hfill \cite{Wang_2023} & -- & \cellcolor{white}2.0E+21 & \cellcolor{white}2.7E+21 & \cellcolor{white}1.3E+22 & \cellcolor{Silver}2.9E+22 & 3.5E+21 & \cellcolor{MyLightGreenYellow}2.2E+21  & \cellcolor{MyLightGreenYellow}1.6E+21 \\
8 & DD$^{\text{(H)}}$* \hfill \cite{Wang_2023} & -- & \cellcolor{white}2.0E+21 & \cellcolor{white}2.4E+21 & \cellcolor{white}1.3E+22 & \cellcolor{Silver}2.9E+22 & 7.0E+21 & \cellcolor{MyYellowOrange}2.3E+21  & \cellcolor{MyYellowOrange}1.6E+21 \\
9 & DT$^{\text{(H)}}$ \hfill \cite{BALBINOT2021100952} & 6.0E+22 & \cellcolor{MyReddishOrange}2.9E+21 & \cellcolor{MyLightGreenYellow}2.3E+22 & \cellcolor{MyGreenish}6.0E+22 & \cellcolor{Silver}1.4E+23 & 5.5E+21 & \cellcolor{MyYellowOrange}1.8E+21  & \cellcolor{MyLightGreenYellow}2.8E+21 \\
10 & ED$^{\text{(H)}}$ \hfill \cite{AHOMANTILA2021100886} & -- & \cellcolor{white}6.5E+23 & \cellcolor{white}4.5E+23 & \cellcolor{white}4.1E+23 & \cellcolor{Silver}9.4E+23 & 5.8E+21 & \cellcolor{MyLightGreenYellow}1.7E+22  & \cellcolor{MyYellowOrange}2.1E+22 \\
11 & ED$^{\text{(H)}}$ \hfill \cite{Subba_2021} & 5.0E+23 & \cellcolor{MyGreenish}6.1E+23 & \cellcolor{MyGreenish}6.4E+23 & \cellcolor{MyGreenish}4.8E+23 & \cellcolor{Silver}1.1E+24 & 1.8E+22 & \cellcolor{MyLightGreenYellow}1.2E+22  & \cellcolor{MyGreenish}2.2E+22 \\
12 & HA$^{\text{(L)}}$* \hfill \cite{Gao_2023} & -- & \cellcolor{white}2.0E+21 & \cellcolor{white}7.6E+20 & \cellcolor{white}-- & \cellcolor{Silver}1.2E+21 & 4.8E+20 & \cellcolor{MyLightGreenYellow}1.6E+20  & \cellcolor{MyGreenish}3.7E+20 \\
13 & JE$^{\text{(L)}}$ \hfill \cite{GROTH2015471} & 8.0E+21 & \cellcolor{MyLightGreenYellow}4.7E+21 & \cellcolor{MyGreenish}6.0E+21 & \cellcolor{white}-- & \cellcolor{Silver}4.4E+22 & -- & \cellcolor{white}2.7E+21  & \cellcolor{white}3.5E+21 \\
14 & JE$^{\text{(L)}}$ \hfill \cite{GROTH2015471} & 9.0E+21 & \cellcolor{MyLightGreenYellow}4.7E+21 & \cellcolor{MyLightGreenYellow}5.6E+21 & \cellcolor{white}-- & \cellcolor{Silver}4.4E+22 & -- & \cellcolor{white}2.7E+21  & \cellcolor{white}3.5E+21 \\
15 & JE$^{\text{(L)}}$ \hfill \cite{GROTH2015471} & 5.0E+21 & \cellcolor{MyLightGreenYellow}8.8E+21 & \cellcolor{MyLightGreenYellow}8.3E+21 & \cellcolor{white}-- & \cellcolor{Silver}4.4E+22 & -- & \cellcolor{white}2.6E+21  & \cellcolor{white}3.5E+21 \\
16 & JE$^{\text{(L)}}$* \hfill \cite{GUILLEMAUT2013S638} & 2.3E+22 & \cellcolor{MyLightGreenYellow}1.5E+22 & \cellcolor{MyLightGreenYellow}1.5E+22 & \cellcolor{white}-- & \cellcolor{Silver}3.9E+22 & -- & \cellcolor{white}1.4E+21  & \cellcolor{white}3.4E+21 \\
17 & J6$^{\text{(H)}}$ \hfill \cite{RUBINO2021100895} & 1.0E+22 & \cellcolor{MyLightGreenYellow}1.6E+22 & \cellcolor{MyLightGreenYellow}2.6E+22 & \cellcolor{MyReddishOrange}7.9E+22 & \cellcolor{Silver}1.8E+23 & 4.2E+21 & \cellcolor{MyGreenish}4.0E+21  & \cellcolor{MyGreenish}5.1E+21 \\
18 & J6$^{\text{(H)}}$ \hfill \cite{RUBINO2021100895} & 2.0E+22 & \cellcolor{MyGreenish}1.6E+22 & \cellcolor{MyLightGreenYellow}3.1E+22 & \cellcolor{MyYellowOrange}7.9E+22 & \cellcolor{Silver}1.8E+23 & 2.8E+21 & \cellcolor{MyGreenish}4.1E+21  & \cellcolor{MyLightGreenYellow}5.1E+21 \\
19 & J6$^{\text{(H)}}$ \hfill \cite{RUBINO2021100895} & 3.0E+22 & \cellcolor{MyLightGreenYellow}1.6E+22 & \cellcolor{MyGreenish}3.7E+22 & \cellcolor{MyLightGreenYellow}7.9E+22 & \cellcolor{Silver}1.8E+23 & 2.8E+21 & \cellcolor{MyLightGreenYellow}4.3E+21  & \cellcolor{MyLightGreenYellow}5.1E+21 \\
20 & J6$^{\text{(H)}}$ \hfill \cite{DeGianni_2024} & 3.6E+22 & \cellcolor{MyLightGreenYellow}1.6E+22 & \cellcolor{MyGreenish}4.5E+22 & \cellcolor{MyLightGreenYellow}7.9E+22 & \cellcolor{Silver}1.8E+23 & -- & \cellcolor{white}4.4E+21  & \cellcolor{white}5.1E+21 \\
21 & MU$^{\text{(H)}}$ \hfill \cite{HAVLICKOVA20151209} & 2.0E+21 & \cellcolor{MyYellowOrange}9.8E+21 & \cellcolor{MyLightGreenYellow}3.5E+21 & \cellcolor{white}-- & \cellcolor{white}-- & 2.1E+23 & \cellcolor{MyReddishOrange}6.5E+21  & \cellcolor{white}-- \\
22 & MU$^{\text{(H)}}$ \hfill \cite{HAVLICKOVA20151209} & 2.0E+21 & \cellcolor{MyYellowOrange}9.8E+21 & \cellcolor{MyGreenish}2.0E+21 & \cellcolor{white}-- & \cellcolor{white}-- & 2.1E+22 & \cellcolor{MyLightGreenYellow}4.8E+22  & \cellcolor{white}-- \\
23 & MU$^{\text{(H)}}$* \hfill \cite{HENDERSON2024101765} & 7.2E+21 & \cellcolor{MyGreenish}8.1E+21 & \cellcolor{white}-- & \cellcolor{MyGreenish}6.4E+21 & \cellcolor{Silver}1.5E+22 & -- & \cellcolor{white}--  & \cellcolor{white}1.1E+21 \\
24 & MU$^{\text{(H)}}$* \hfill \cite{HENDERSON2024101765} & 1.0E+22 & \cellcolor{MyGreenish}8.1E+21 & \cellcolor{white}-- & \cellcolor{MyLightGreenYellow}6.4E+21 & \cellcolor{Silver}1.5E+22 & -- & \cellcolor{white}--  & \cellcolor{white}1.1E+21 \\
25 & MU$^{\text{(L)}}$* \hfill \cite{Moulton_2024} & 4.0E+21 & \cellcolor{MyLightGreenYellow}7.1E+21 & \cellcolor{MyLightGreenYellow}2.2E+21 & \cellcolor{white}-- & \cellcolor{Silver}6.0E+21 & -- & \cellcolor{white}3.6E+21  & \cellcolor{white}7.5E+20 \\
26 & MU$^{\text{(L)}}$* \hfill \cite{Moulton_2024} & 5.0E+20 & \cellcolor{MyReddishOrange}1.7E+22 & \cellcolor{MyLightGreenYellow}1.4E+21 & \cellcolor{white}-- & \cellcolor{Silver}6.0E+21 & -- & \cellcolor{white}3.3E+22  & \cellcolor{white}7.5E+20 \\
27 & MU$^{\text{(H)}}$* \hfill \cite{Verhaegh_2023} & 1.5E+21 & \cellcolor{MyReddishOrange}9.8E+21 & \cellcolor{white}-- & \cellcolor{MyYellowOrange}5.2E+21 & \cellcolor{Silver}1.2E+22 & -- & \cellcolor{white}--  & \cellcolor{white}9.3E+20 \\
28 & MU$^{\text{(H)}}$* \hfill \cite{Verhaegh_2023} & 6.0E+21 & \cellcolor{MyLightGreenYellow}9.8E+21 & \cellcolor{white}-- & \cellcolor{MyLightGreenYellow}3.5E+21 & \cellcolor{Silver}8.2E+21 & -- & \cellcolor{white}--  & \cellcolor{white}8.3E+20 \\
29 & MU$^{\text{(H)}}$ \hfill \cite{Myatra_2023} & 3.2E+21 & \cellcolor{MyYellowOrange}9.8E+21 & \cellcolor{MyLightGreenYellow}5.3E+21 & \cellcolor{MyGreenish}2.4E+21 & \cellcolor{Silver}5.6E+21 & -- & \cellcolor{white}4.9E+21  & \cellcolor{white}7.3E+20 \\
30 & MU$^{\text{(H)}}$ \hfill \cite{Myatra_2023} & 2.0E+21 & \cellcolor{MyYellowOrange}9.8E+21 & \cellcolor{MyLightGreenYellow}4.5E+21 & \cellcolor{MyGreenish}2.4E+21 & \cellcolor{Silver}5.6E+21 & 7.0E+20 & \cellcolor{MyReddishOrange}5.3E+21  & \cellcolor{MyGreenish}7.3E+20 \\
31 & MP* \hfill \cite{Tanaka_2020} & 1.5E+22 & \cellcolor{MyGreenish}1.8E+22 & \cellcolor{white}-- & \cellcolor{white}-- & \cellcolor{white}-- & -- & \cellcolor{white}--  & \cellcolor{white}-- \\
32 & NU \hfill \cite{Chen_2018} & 7.8E+21 & \cellcolor{MyReddishOrange}6.0E+20 & \cellcolor{MyLightGreenYellow}4.4E+21 & \cellcolor{white}-- & \cellcolor{Silver}4.3E+22 & -- & \cellcolor{white}3.6E+21  & \cellcolor{white}1.5E+21 \\
33 & NU \hfill \cite{Chen_2018} & 6.2E+21 & \cellcolor{MyReddishOrange}6.0E+20 & \cellcolor{MyGreenish}4.4E+21 & \cellcolor{white}-- & \cellcolor{Silver}4.3E+22 & -- & \cellcolor{white}3.6E+21  & \cellcolor{white}1.5E+21 \\
34 & SP$^{\text{(H)}}$ \hfill \cite{Lore_2024} & 3.0E+22 & \cellcolor{MyReddishOrange}3.1E+21 & \cellcolor{MyLightGreenYellow}1.6E+22 & \cellcolor{MyLightGreenYellow}5.3E+22 & \cellcolor{Silver}1.2E+23 & 1.4E+21 & \cellcolor{MyGreenish}1.3E+21  & \cellcolor{MyGreenish}2.0E+21 \\
35 & TC$^{\text{(L)}}$* \hfill \cite{HARRISON20171071} & 2.2E+21 & \cellcolor{MyLightGreenYellow}7.5E+20 & \cellcolor{MyGreenish}1.6E+21 & \cellcolor{white}-- & \cellcolor{Silver}2.3E+21 & -- & \cellcolor{white}4.8E+20  & \cellcolor{white}2.7E+20 \\
36 & TC$^{\text{(H)}}$ \hfill \cite{Yang2023Control} & 6.0E+20 & \cellcolor{MyLightGreenYellow}1.5E+21 & \cellcolor{MyLightGreenYellow}1.6E+21 & \cellcolor{MyLightGreenYellow}1.1E+21 & \cellcolor{Silver}2.6E+21 & -- & \cellcolor{white}3.2E+20  & \cellcolor{white}2.8E+20 \\
37 & WE* \hfill \cite{RIVALS2024101723} & 1.2E+20 & \cellcolor{MyLightGreenYellow}3.0E+20 & \cellcolor{MyLightGreenYellow}1.8E+20 & \cellcolor{white}-- & \cellcolor{Silver}1.9E+22 & 6.1E+21 & \cellcolor{MyReddishOrange}2.6E+20  & \cellcolor{MyYellowOrange}1.1E+21 \\
38 & WE$^{\text{(L)}}$ \hfill \cite{YANG2022101302} & 1.0E+20 & \cellcolor{MyLightGreenYellow}3.0E+20 & \cellcolor{MyYellowOrange}3.8E+20 & \cellcolor{white}-- & \cellcolor{Silver}4.4E+21 & -- & \cellcolor{white}1.2E+20  & \cellcolor{white}5.4E+20 \\
39 & WE$^{\text{(L)}}$ \hfill \cite{YANG2022101302} & 1.1E+20 & \cellcolor{MyLightGreenYellow}3.0E+20 & \cellcolor{MyYellowOrange}3.8E+20 & \cellcolor{white}-- & \cellcolor{Silver}4.4E+21 & -- & \cellcolor{white}1.2E+20  & \cellcolor{white}5.4E+20 \\
40 & WE$^{\text{(L)}}$ \hfill \cite{YANG2022101302} & 2.8E+20 & \cellcolor{MyGreenish}3.0E+20 & \cellcolor{MyGreenish}3.6E+20 & \cellcolor{white}-- & \cellcolor{Silver}4.4E+21 & -- & \cellcolor{white}1.2E+20  & \cellcolor{white}5.4E+20 \\
41 & WE$^{\text{(L)}}$ \hfill \cite{YANG2022101302} & 3.5E+20 & \cellcolor{MyGreenish}3.0E+20 & \cellcolor{MyGreenish}3.5E+20 & \cellcolor{white}-- & \cellcolor{Silver}4.4E+21 & -- & \cellcolor{white}1.2E+20  & \cellcolor{white}5.4E+20 \\
42 & WE$^{\text{(L)}}$ \hfill \cite{YANG2022101302} & 4.9E+20 & \cellcolor{MyLightGreenYellow}3.0E+20 & \cellcolor{MyGreenish}3.4E+20 & \cellcolor{white}-- & \cellcolor{Silver}4.4E+21 & -- & \cellcolor{white}1.2E+20  & \cellcolor{white}5.4E+20 \\
43 & WE$^{\text{(L)}}$ \hfill \cite{YANG2022101302} & 7.6E+20 & \cellcolor{MyLightGreenYellow}3.0E+20 & \cellcolor{MyLightGreenYellow}3.1E+20 & \cellcolor{white}-- & \cellcolor{Silver}4.4E+21 & -- & \cellcolor{white}1.1E+20  & \cellcolor{white}5.4E+20 \\
44 & WE$^{\text{(L)}}$ \hfill \cite{Yang_2024} & 9.7E+20 & \cellcolor{MyYellowOrange}3.0E+20 & \cellcolor{MyLightGreenYellow}3.7E+20 & \cellcolor{white}-- & \cellcolor{Silver}5.3E+21 & -- & \cellcolor{white}1.2E+20  & \cellcolor{white}5.8E+20 \\
45 & WE$^{\text{(L)}}$ \hfill \cite{Yang2023Control} & 5.0E+20 & \cellcolor{MyLightGreenYellow}3.0E+20 & \cellcolor{MyGreenish}3.4E+20 & \cellcolor{white}-- & \cellcolor{Silver}4.9E+21 & -- & \cellcolor{white}1.2E+20  & \cellcolor{white}5.6E+20 \\
\midrule
\multicolumn{2}{c}{Avg. agreement:} & -- & \cellcolor{MyYellowOrange}4.2 (2.6) & \cellcolor{MyLightGreenYellow}1.9 (1.7) & \cellcolor{MyLightGreenYellow}2.3 (1.9) & -- & -- & \cellcolor{MyYellowOrange}5.6 (2.9) & \cellcolor{MyLightGreenYellow}2.1 (1.8)\\
\bottomrule
\end{tabular}

%% file: 2-table-examples.tex
\begin{tabular}{lcccccc}
\toprule
  & ID & ARC \cite{SORBOM2015378_ARC}
 & MANTA \cite{Rutherford_2024}
 & Infinity Two* \cite{Bader}
 & Stellaris* \cite{LION2025114868}
 & FNSF \cite{Islam_2024} \\
\midrule
Actual $\Gamma_{\text{D}}$ & --                              & --        & --        & --        & --        & 4.0E+22 \\
Eq. (\ref{eq:Vofl_vs_Qpuff}) & VOL             & 1.2E+23   & 1.1E+23   & 3.2E+22   & 9.5E+21   & 1.0E+22 \\
Eq. (\ref{eq:V_OFL_S_OFL_1p5_times_opacity_vs_Qpuff}) & OPQ
                                                 & 1.5E+23   & 1.4E+23   & 1.0E+23   & 2.1E+21   & 4.8E+22 \\
Eq. (\ref{eq:HDL_tokamaks}) & HDL              & --        & --        & --        & --        & 1.9E+22 \\
Eq. (\ref{eq:V_OFL_S_OFL_1p5_times_opacity_vs_Qpuff_tokamaks}/\ref{eq:V_OFL_S_OFL_1p5_times_opacity_vs_Qpuff_stellarators}*) & GES/Su*
                                                 & 1.7E+23   & 1.5E+23   & 1.3E+23   & 2.2E+23   & 4.4E+22 \\\midrule
Actual $\Gamma_{\text{Z}}$ & --                              & --        & --        & --        & --        & 1.0E+21 \\
Eq. (\ref{eq:GammaZ_final}) & via $\fQZ$        & 7.7E+21   & 2.2E+21   & 9.5E+20   & 5.4E+20   & 2.7E+21 \\
Eq. (\ref{eq:Zpuff_tokamaks_GW}/\ref{eq:Zpuff_stellarators_SU}*) & GES/Su*
                                                 & 4.8E+21   & 4.9E+21   & 5.2E+21   & 6.3E+21   & 3.4E+21 \\
\bottomrule
\end{tabular}

%% file: 3-table-appendix.tex
\begin{tabular}{llll}
\toprule
\# & Machine & ID & References \\
\midrule
1 & Alcator C-Mod & AM & \cite{LIPSCHULTZ1997771, REKSOATMODJO2021100971, Reinke_2019, LORE2015515, Lore_2015, Goetz_1999} \\
2 & AUG & AU & \cite{Henderson_2023, Kallenbach_2018, Lunt_PRL_CRD, Senichenkov_2021, Potzel_2014, Kallenbach_2015, Kallenbach_2013, Brida_2017, WU2023114023, Wu_2021, SYTOVA201972, Senichenkov_2019, SENICHENKOV2023101361, Reimold_2015, Ivanova_Stanik_2022, pan_2024_thesis, CARRALERO20171189, Bernert_2021, BERNERT2017111, Kallenbach_2021} \\
3 & CFETR & CF & \cite{MAO20151233} \\
4 & COMPASS & CO & \cite{Komm_2019, Komm_2021, Dimitrova_2020} \\
5 & DIII-D & DD & \cite{Lan_2020, Mordijck_2020, Wang_2023, PETRIE1992848, PETRIE2007416, Casali_2022, PETRIE2011S906, McKee_2000, Wu_2024, Wang2021} \\
6 & DT-ALPHA & DA & \cite{YOSHIMURA20251401015} \\
7 & DTT & DT & \cite{BALBINOT2021100952, Moscheni_2022, Moscheni_2025, INNOCENTE2021100985} \\
8 & EAST & EA & \cite{DING2022101250, Wang_2019, ELDON2021100963} \\
9 & EU-DEMO & ED & \cite{ZAGORSKI201637, AHOMANTILA2021100886, Subba_2021, Korzueva_2023, Xiang_2021} \\
10 & FNSF & FN & \cite{Islam_2024, ISLAM2022101292} \\
11 & Globus-M2 & GM & \cite{sorokina2018testing} \\
12 & HL-2A & HA & \cite{Gao_2023} \\
13 & HL-2M & HM & \cite{ZHOU2022113222, Zhang_2022} \\
14 & ITER & IT & \cite{Lore_2022, VESELOVA2021100870, SURESHKUMAR2024101780, Kaveeva_2020} \\
15 & JET & JE & \cite{GUILLEMAUT2013S638, Telesca_2022, Loarte_1998, GROTH2015471, Glöggler_2019, TELESCA2017882, TELESCA2015577, maddison2009impurity, huber2014impact, ZAGORSKI2015649, MADDISON2011S313, Huber_2020} \\
16 & JFT-2M & J2 & \cite{Kawashima_1999} \\
17 & JT-60SA & J6 & \cite{RUBINO2021100895, DeGianni_2024, Yamoto_2023} \\
18 & KSTAR & KS & \cite{gupta2025detachmentcontrolkstartungsten} \\
19 & LHD & LH & \cite{MUKAI2022101294, MASUZAKI2013S133, miyazawa2005self} \\
20 & Magnum-PSI & MP & \cite{Tanaka_2020} \\
21 & MAP-II & MI & \cite{Okamoto_2006} \\
22 & MAST-U & MU & \cite{HAVLICKOVA20151209, Verhaegh_2023, HENDERSON2024101765, Moulton_2024, Myatra2021Numerical, Verhaegh_2021} \\
23 & NSTX & NS & \cite{MEIER20151200, Soukhanovskii_2009, soukhanovskii2006divertor, Soukhanovskii_2009_2} \\
24 & NSTX-U & NU & \cite{Chen_2018} \\
25 & Pilot-PSI & PP & \cite{Hayashi_2016} \\
26 & SPARC & SP & \cite{Lore_2024} \\
27 & STEP & ST & \cite{Osawa_2024, Osawa_2023} \\
28 & TCV & TC & \cite{FEVRIER2021100977, Fil_2020, HARRISON20171071, Yang2023Control, Yang_2023, Février_2020, Harrison_2019, Mombelli_2025, theiler2018sol, Smolders_2020, Stagni_2022} \\
29 & TPD-Sheet IV & TS & \cite{Onda_2017, takimoto2017experimental} \\
30 & W7-AS & WA & \cite{McCormick_2002_PRL} \\
31 & W7-X & WX & \cite{Effenberg_2019, effenberg2018demonstration} \\
32 & WEST & WE & \cite{Yang_2024, RIVALS2024101723, YANG2022101302, maget2021nitrogen} \\
\bottomrule
\end{tabular}